\documentstyle[12pt]{article}
\input epsf
\begin{document}

\title{Interplay of Hard and Soft Physics in Small $x$ Deep Inelastic
  Processes}

\author{Halina Abramowicz\thanks{\noindent Supported by GIF project No
    127402207/93}\\ School of Physics and Astronomy\\ Raymond and
  Beverly Sackler Faculty of Exact Sciences\\ Tel Aviv
  University
\\[0.4cm]
Leonid Frankfurt\thanks{\noindent On leave of
    absence from the St.Petersburg Nuclear Physics Institute, Russia.
    Supported by BSF Grant No. 9200126 }\\ School of Physics and
  Astronomy\\ Raymond and Beverly Sackler Faculty of Exact Sciences\\
  Tel Aviv University
\\[0.4cm]
and \\[0.4cm]
Mark Strikman\thanks{\noindent Also  St.Petersburg Nuclear
Physics Institute, Russia. Supported by DOE Contract  DE-FG02-93ER40771 }\\
Pennsylvania State  University, University Park
}

\maketitle

\vspace{0.7cm}

\vspace{-20cm}
\begin{flushleft}
\tt DESY 95-047 \\
March 1995 \\
\end{flushleft}

\newpage

\begin{abstract}

Coherence phenomena, the increase with energy of coherence length and the
non-universality of parton structure of the effective Pomeron are explained.
New hard phenomena directly calculable in QCD such as diffractive
electroproduction of states with $M^2\ll Q^2$ and the color transparency
phenomenon as well as new options to measure the light-cone wave functions of
various hadrons are considered. An analogue of Bjorken scaling is predicted
for the diffractive electroproduction of $\rho$ mesons at large momentum
transfers and for the production of large rapidity gap events, as observed at
HERA.  A phenomenological QCD evolution equation is suggested to calculate the
basic characteristics of the large rapidity gap events. The increase of parton
densities at small $x$ as well as new means to disentangle experimentally soft
and hard physics are considered. We discuss constraints on the increase of
deep inelastic amplitudes with $Q^2$ derived from the inconsistency of QCD
predictions for inclusive and exclusive processes and from unitarity of the S
matrix for collisions of wave packets. New ways to probe QCD physics of hard
processes at large longitudinal distances and to answer the long standing
problems on the origin of the Pomeron are suggested.  Unresolved problems and
perspectives of small $x$ physics are also outlined.

\end{abstract}

\section{Introduction}

The aim of this report is to outline QCD predictions for color
coherence phenomena -- a result of nontrivial interplay of hard and
soft QCD physics specific for high energy processes. Coherence
phenomena provide an important link between the well understood
physics of hard processes and the physics of soft processes which at
present is mostly phenomenological.  The soft/hard interplay is
elaborated for the exclusive deep inelastic processes $\gamma^*_L + N
\rightarrow a +N$ for $M_a^2 \ll Q^2$ directly calculable in QCD.
These processes provide new methods of investigating the structure of
hadrons and the origin of the Pomeron and allow to search for new
forms of hadronic matter in heavy ion collisions (for a review and
references see~\cite{FMS94}). The phenomenon of coherence reveals
itself in high energy processes through a large probability of
occurrence of diffractive processes and through their specific
properties.  Thus in this report we concentrate mostly on diffractive
processes. To explain the role of coherence in high energy processes
we will consider some properties of QCD which are difficult to
reconcile with intuition based on pre-QCD ideas and on perturbative QCD
(PQCD) experience with medium $x$ processes.

Let us outline briefly the experimental results pointing to
a significant role of color coherent effects. \\
(i) Observed fast increase of parton distributions at small $x$ and
large momentum transfer $Q$~\cite{ZEUSa,H1a}, which in the case of the
ZEUS measurements can be parameterized as\footnote{This
  parameterization is not applicable for $Q^2\geq 10^4$~GeV$^2$
  where it leads to a divergence of the momentum sum rule which is
  precise in QCD.}
\begin{equation}
  F_{2p}(x,Q^2)= (1-x^2)^4 \left[ 0.35+0.017 x^ {-(.35+.16\log_{10}
    Q^2)} \right] \ .
\label{eq:lc}
\end{equation}
(ii) Large cross section measured for leptoproduction of vector mesons
at CERN~\cite{NMC1} and HERA~\cite{ZEUSb} which increases with energy
as
\begin{equation}
\sigma(\gamma^* + p \rightarrow \rho + p) \propto \sigma_{tot}^2(\gamma^*
+ p ) \ .
\label{eq:2c}
\end{equation}
The energy dependence obtained by the  HERA data is much
faster than that of cross sections for soft two body hadron processes
like elastic $pp$ collisions.  At the same time this observation is in
line with the QCD predictions for the electroproduction of vector
mesons and for
color transparency phenomenon which are described in sections 4-6.

\noindent (iii) Large probability of occurrence of large rapidity gap
events observed at HERA in the deep inelastic
regime~\cite{ZEUSgap,H1gap} -- this is another evidence for the
important role of soft physics in deep inelastic small $x$ processes.
HERA data ~\cite{MD} support the dominance of events with low
transverse momenta $k_t$.  A significant magnitude of nuclear
shadowing in nucleus structure functions observed at CERN and at FNAL
is yet another evidence for the important role of soft physics in the
deep inelastic processes at small $x$ (for a recent review of
experimental data see~\cite{Arneodo} and for a theoretical discussion
of nuclear shadowing see~\cite{FS88,FS91} and references therein).

\noindent (iv) Presence of significant fluctuations in the strength of
high--energy hadron--nucleon interactions as indicated by the analysis
of soft diffractive processes with protons and nuclei. In particular,
the observed cross sections for inelastic diffractive production of
states X off nuclei~\cite{Ferbel}, $h + A \rightarrow X + A$,
increases with the atomic number $A$ significantly faster than the
expected $A^{{1\over 3}}$ if hadrons were to interact with a nucleus
as with a black body.

\section{ Glossary and notation.}

In the course of the presentation we will use a customary notation for
variables used in describing deep inelastic phenomena. For convenience
we define them below:

\noindent -- $q$ denotes the 4--momentum vector of a virtual photon
\[ q = (q_0, \vec{q}) \]
and \[ Q^2=-q^2. \]

\noindent -- $x$ is the Bjorken variable
\[ x=\frac{Q^2}{2pq} \ ,\]
where $p$ denotes the 4--momentum vector of the target.
In the case of nuclear targets with atomic number $A$,
$x$ is defined as \[ x=\frac{A~Q^2}{2pq} \ .\]

\noindent -- $m_N$ stands for the mass of the nucleon.

\noindent -- $s$ denotes the square of the center of mass energy
available in the collision of a projectile with a target.

For the sake of simplicity we do not always explain all the variables
if the notation is self explanatory. Thus we use $E$ to denote the
energy of a state or a particle, $M$ for the mass of a state and $m$ for
the mass of a particle. The subscripts explain further the objects
considered.

We will try to consistently denote by $N$ a nucleon target, by $A$ a
nuclear target and by $T$ any target ($N$ or $A$).

In the splitting of the photon into a $q \bar q$ we will denote by $z$
the fraction of the photon momentum carried by one of the quarks of
the pair and by $k_t$ their transverse momentum relative to the photon
direction.

When we refer to the small or low $x$ region we have in mind $x$
typically less than
 $10^{-2}$ which corresponds roughly to the value of $x$ where one
starts observing a rise of parton densities in the proton.
Medium $x$ refers to $10^{-2}<x<10^{-1}$.

\section{ Increase of coherence length with energy.}

The starting point of our discussion is the pre-QCD
suggestion~\cite{Gribov65} that distances along a projectile momentum
direction, $l_c$, which contribute to high energy processes increase
with energy. The underlying nontrivial physical picture is that a
sufficiently energetic projectile transforms into a hadronic component
at longitudinal distances $l_c$ from the target which are large and
increase with energy. In the case of deep inelastic processes the
typical longitudinal distances, in the target rest frame, are
described at small $x$ by formula~\cite {BJoffe68}
\begin{equation}
l_c={1\over 2m_N x} \ .
\label{eq:3c}
\end{equation}
This formula can be also understood as a consequence of the
uncertainty principle and the renormalizability of QCD. The life time
$\tau$ of a virtual photon with momentum $q$ in a hadron configuration
$|n\rangle$ with mass $M_{n}$ is given by the uncertainty principle as
\begin{equation}
\tau ={1 \over E_{n}-q_{o}}\simeq {2q\over (M_n^2+Q^2)}
\simeq {1\over 2m_N x} \ .
\label{eq:4c}
\end{equation}
In the last step of the derivation of the above estimate, as a
consequence of QCD renormalizability, the contribution of masses
$M_n^2\gg Q^2$ can been neglected and $M_n^2$ can been approximated by
$Q^2$.

In PQCD calculations of small $x$ processes the increase with energy
of $l_c$ is a direct consequence of gauge invariance and the
renormalizability of PQCD.  In the nonperturbative QCD regime within
the parton model, equation~(\ref{eq:4c}) follows from fast convergence
of the integrals over transverse momenta of constituents.

We now turn to diffractive production of states $X$ with mass $M_X$ in
the reactions $a + T \rightarrow X +T$ where $a$ is a fast projectile
and $T$ the target. For the sake of the argument we will assume that
$a$ is a photon with virtuality $Q^2$.  The minimal momentum
transferred to the target in such a process is given by
\begin{equation}
t_{min} =-{(M_{X}^2+Q^2)^2 \over s^2} m_{T}^2 \ .
\label{eq:5c}
\end{equation}
where $s=(q+p_{T})^2$
is the square of the center of mass energy  of
the $\gamma^* T$ collision.
If at sufficiently large $s$
\begin{equation}
-{t_{min} r_{T}^2\over 3} \ll 1,
\label{eq:6c}
\end{equation}
the state $X$ can be produced without disturbing the target, that is
without form factor suppression. Here $r_T^2$ is the average quadratic
radius of the target. Thus with increasing $s$ the variety of
diffractive processes increases. At the same time the number of
possible intermediate parton configurations in the wave function of
the photon which do not destroy the coherence of the target in the
interaction also increases. In the calculation of the diffractive
cross section the contribution of all these intermediate states has to
be summed coherently~\footnote {For soft hadron processes this is
  implemented in the Gribov Reggeon Calculus \cite{Gribov4}.}.  For
some hard diffractive processes the sum over hadron states can be
expressed through the parton distributions in the target (sections 3
and 4).  The Fourier transform of the forward diffractive amplitude
into coordinate space shows that the dominant contributions come from
longitudinal distances concentrated around $l \sim {1\over
  \sqrt{-t_{min}}}$.  This is essentially the same result as
established in equation~(\ref{eq:4c}).

{}From all the above considerations one can conclude that the mechanism
of coherence is related to a large probability of coherent diffractive
processes and that $l_{c}$ is a measure of the coherence length.

The validity of the formula given by equation~(\ref{eq:4c}) has been
established by B.Ioffe~\cite{BJoffe68} in an analysis of the Fourier
transform of data on structure functions.  Since the same distances
are important in vacuum and non-vacuum channels~\cite{FS88}, another
evidence for the validity of equation~(\ref{eq:4c}) is the energy
dependence and the significant value of nuclear shadowing in
photoproduction and in the deep inelastic processes at large energies
(for a recent review and references see~\cite{FS88,FS91}).  To be more
quantitative, let's consider a simple example which will be important
for further considerations in this report.  In the deep inelastic
processes at HERA at $x=10^{-4}$ the value of $l_c$ in the proton rest
frame comes out to be
\begin{equation}
l_c = 10^3~{\rm fm} \ .
\label{eq:7c}
\end{equation}
This is a macroscopic distance on the scale of hadronic physics.
In the simplest case of a virtual photon fluctuating into a pair of
almost free quarks, the pair can propagate macroscopic distances
without confinement.  At LHC energies in high $p_t$ phenomena, $l_c$
may achieve atomic sizes -- $10^5 {\rm~fm}$.
The longitudinal size of the fast $q \bar q$ state is
 rather small, $\sim {1 \over Q}$. However, since the $q \bar q$ pair
 is not an eigenstate of the QCD Hamiltonian, radiation of quark and
 gluons will occur during the space-time evolution of the wave package
 and its longitudinal size may reach 1 fm by the time it hits the
 target.

Thus the space time description of small $x$ processes in the target rest
frame leads to the conclusion that the
quark-gluon configuration of the fast projectile involved in the
collision is build
 over macroscopic distances (on the scale of hadron physics).
It is this property of small $x$ physics which will lead to many
coherent phenomena.
  In this report we will often use the target rest frame when
discussing coherence effects. Since the results cannot be frame
dependent, the same effects are present in the more habitual infinite
momentum reference frame. The choice of frame is thus a matter of
convenience.  We will show that in the infinite momentum frame the
same picture leads to short range correlations in rapidity space for
small $x$ partons. This will be important for the description of large
rapidity gap events.

\section{ Interaction  cross section for small size wave packet.}

One of the striking QCD predictions for hard processes dominated by
large longitudinal distances is that if a hadron is found in a small
size configuration of partons it interacts with a target with a small
cross section. This prediction which follows from the factorization
theorem for hard processes in QCD is in variance with many
phenomenological approaches based on pre-QCD ideas and on quark models
of hadrons.

A sufficiently energetic wave packet with zero baryon and color
charges localized in a small transverse volume in the impact parameter
space can be described by a $q\bar q$ pair .  This conclusion follows
from asymptotic freedom in QCD which implies that the contribution of
other components is suppressed by a power of the strong coupling
constant $\alpha_s$ and/or a power of $Q^2$.  A familiar example of
such a wave packet is a highly virtual longitudinally polarized
$\gamma^*$ in a $q\bar q$ state.  Within the parton model the cross
section for the interaction of such a photon with a target is
suppressed by a power of $Q^2$.  But at the same time the probability
for a longitudinal photon to be in a large transverse size
configuration (soft physics=parton model contribution) is suppressed
by a power of $Q^2$.  These properties explain why reactions initiated
by longitudinally polarized photons are best to search for new QCD
phenomena.

The cross section for a high-energy interaction of a small size $q\bar
q$ configuration off any target can be unambiguously calculated in QCD
for low $x$ processes by applying the QCD factorization theorem.  In the
approximation when the leading
 $\alpha_s \ln {Q^2\over \Lambda_{QCD}^2}
\ln x$ terms are accounted for~\cite{BBFS93,FMS93} the
result is
\begin{equation}
\sigma(b^2)={\pi^2\over 3}\left[ b^2\alpha_s(Q^2)xG_{T}
(x,Q^2)\right]_{x=Q^2/s,Q^2 \simeq 15/b^2} \ ,
\label{eq:9c}
\end{equation}
where $b$ is the transverse distance between the quark $q$ and the
antiquark $\bar q$ and $G_T(x,Q^2)$ is the gluon distribution in the
target $T$ calculated within this approximation. In this equation the $Q^2$
evolution and the small $x$
physics are properly taken into account through the gluon
distribution.

It is possible to derive similar  equation in
the  leading  $\alpha_s \ln {Q^2\over \Lambda_{QCD}^2}$ approximation one
should account for all hard processes including diagrams where (anti)quarks in
the box diagram exchange gluon. The final result has the same form as
eq.(\ref{eq:9c}), but with $G_N(x,Q^2)$ calculated in
 the  leading  $\alpha_s \ln {Q^2\over \Lambda_{QCD}^2}$ approximation. It also
contains a
small contribution due to sea quarks.
Eq.(\ref{eq:9c}) accounts for the contribution of quarks $Q$ whose masses
satisfy the condition: $l_c = {2 q_0 \over 4 m_Q^2} \gg
r_N$
The estimate $Q^2 \approx {15 \over b^2}$ was obtained
in \cite{FKS} by numerical analysis
of the $b$-space representation of the
cross section of the longitudinally polarized photon, $\sigma_L$,
 and requiring
that $G_T$ is conventional gluon
distribution calculated in the leading
 $\alpha_s \ln {Q^2\over \Lambda_{QCD}^2}$ approximation.

The generalization of equation~(\ref{eq:9c}) for
interactions of small size wave packets with nonzero baryon number is
straightforward
but technically rather cumbersome~\cite {FRS95}.

There is a certain similarity between
equation~(\ref{eq:9c}) and the two gluon exchange model of
F. Low~\cite{Low} and S. Nussinov~\cite{Nussinov}, as well as the
constituent quark 2 gluon exchange model of J. Gunion and D.
Soper~\cite{GS}.  The factor $b^2$ which is present in the QCD
expression~(\ref{eq:9c}) for the cross section is also present in
these models.  The major distinction between the results of QCD
calculations and the two gluon exchange models is the presence of
terms involving both the gluon and the sea quark distributions
 in equation~(\ref{eq:9c}).  The latter are  particularly
relevant for the fast increase of the cross section at small $x$,
for the seemingly slow decrease with $Q^2$ of higher twist processes
and for the increase of nuclear shadowing with decreasing $x$. All
those effects are characteristic for QCD as a gauge quantum field
theory which predicts an increase of parton densities in hadrons with
$\frac{1}{x}$
in contrast to quantum mechanical models of hadrons.
Another salient property of QCD as a renormalizable quantum field
theory is that for hard processes the cross sections have to be
expressed in terms of parton distributions in a target, where partons
are the bare particles of the QCD Lagrangian. This is not the case in
the quantum mechanical models of hadrons used in~\cite{NZ,NZZ} where
hard processes are modeled in terms of constituent quarks interacting
through the exchange of gluons with non--zero mass. The attempt to
reinterpret the gluon distribution in equation~(\ref{eq:9c}) as due to
Weizs\"acker-Williams gluons arising from the color field of constituent
quarks~\cite{NZZ} is in variance with the QCD evolution equation
analysis of the measured structure functions~\cite{MRS,CTEQ,GRV}. Such
analysis points to an important role of the valence sea and gluons in
the nonperturbative parton distributions of a hadronic
target even at a very low normalization point~\cite{GRV,MRSLQ}.

In QCD the inelastic cross section for the collision of a sufficiently
energetic small size, colorless two gluon configuration off any target
is,
\begin{equation}
\sigma(b^2)={3\pi^2\over 4}\left[ b^2\alpha_s(Q^2)xG_{T}
(x,Q^2)\right]_{x=Q^2/s, Q^2=\lambda/b^2} \ ,
\label{eq:9d}
\end{equation}
where the parameter $\lambda$ is likely to be similar to the one
present in the case of scattering of a $q \bar q$ pair off a target.
The difference compared to equation~(\ref{eq:9c}) is in the factor $9/4$
which follows from the fact that gluons belong to the octet representation
of the color group $SU(3)_c$ while quarks are color triplets.

\section{Electroproduction of vector mesons in QCD.}

One of the examples of a new kind of hard processes calculable in QCD
is the coherent electroproduction of vector mesons off a target T,
\begin{equation}
\gamma^* + T \rightarrow V+T \ ,
\label{eq:10c}
\end{equation}
where $V$ denotes any vector meson ($\rho,\omega,\phi, J/\Psi$) or
its excited states.

The idea behind the calculation of hard diffractive processes is that
when $l_{c}$ given by equation~(\ref{eq:4c}) exceeds the diameter of
the target, the virtual photon transforms into a hadron component well
before reaching the target and the final vector meson $V$ is formed
well past the target. The hadronic configuration of the final state is
a result of a coherent superposition of all those hadronic
fluctuations of the photon that satisfy equation~(\ref{eq:6c}).  Thus,
as in the more familiar leading twist deep inelastic processes, the
calculation should take into account all possible hadronic
intermediate states satisfying~equation~(\ref{eq:6c}).  The use of
completeness over diffractively produced intermediate hadronic states
allows to express the result in terms of quarks and gluons as in the
case of other hard processes.  The matrix element of electroproduction
of a vector meson $A$ can be written as a convolution of the light cone
wave function of the photon $\psi^{\gamma^* \rightarrow |n\rangle}$ ,
the scattering amplitude for the hadron state $|n \rangle$,$~A (nT)$,
and the wave function of the vector meson $\psi_{V}$
\begin{equation}
A= \psi^{* ~\gamma^*  \rightarrow |n\rangle}  \otimes
A (nT) \otimes \psi_{V} \ .
\label{eq:11c}
\end{equation}
In the case of a longitudinally polarized photon with high $Q^2$ the
intermediate state $|n\rangle$  is  a $q\bar q$ pair. As was
mentioned in the previous chapter, it  can be demonstrated
by direct calculations that the  contribution of higher Fock state
components  and soft physics are suppressed by a factor $1 \over Q^2$
and/or powers of $\alpha_s$. The proof of this result resembles
the calculation of the total cross section for the deep inelastic scattering
in QCD. The situation is qualitatively different in  the case of a
transversely polarized photon due to the singular behavior of the vertex
$\gamma^*_T \rightarrow q\bar q$ when one of the partons carries a small
fraction of the photon momentum. In this case  soft
and hard physics compete in a wide range of $Q^2$ (see discussion
in  sections 10 and 11).

To understand the applicability of PQCD for the process discussed
above it is convenient to perform the Fourier transform of the
amplitude into the impact parameter space which leads to
\begin{equation}
 A \propto Q\int b^2 x G_T(x,b^2)
K_0\left(Qb\sqrt{z(1-z)}\right)
\psi_{V}(z,b) d^2\!b
z(1-z) dz \ ,
\label{brep}
\end{equation}
where $z$ denotes the fraction of the photon momentum carried by one
of the quarks. Here
\begin{equation}
\psi^{\gamma^*_{L}}\propto z(1-z)Q
K_0\left(Qb\sqrt{z(1-z)}\right) \ ,
\end{equation}
where $K_0$  is the Hankel function of an imaginary argument.
To estimate which values of $b$ dominate in the integral we
approximate $\psi_V(z,k_t)$ by ${z(1-z) \over (k_t^2+\mu^2)^2}$ which
corresponds to $\psi_{V}(z,b)\propto  z(1-z)b K_1(\mu b)$.
We vary  $\left< k^2_t\right>^{1/2} = {\mu \over \sqrt{2}}$
 between 300 and 600 MeV/c.

 We find that in case of $\sigma_L$ the average transverse size
$\left< b \right> \simeq 0.25 {\rm~fm} $ for $Q^2 = 10 {\rm~GeV}^2, x \sim
10^{-3}$ and decreases at larger $Q^2$ approximately as
$ 0.3 fm{3 GeV \over Q}$. It also weakly decreases with decreasing
   $x$.  The increase of $G_T(x,b^2)$, in equation~(\ref{brep}) with
   decreasing $b$ substantially contributes to the decrease of $\left< b
   \right>$.  In the case of a transversely polarized $\gamma^*$ the
   contribution of large $b$ is not suppressed since
\begin{equation}
  \psi^{\gamma^*_{T}}\propto {\partial \over \partial b_{\mu}}
K_0\left(Qb\sqrt{z(1-z)}\right) \ .
\end{equation}
and therefore the contribution of the kinematical region $z
\rightarrow 0$ and $z \rightarrow 1$ where nonperturbative QCD
dominates is not suppressed.

It is worth noting that $\left< b \right>$ contributing in the
calculation of $\sigma_L$ -- $\left< b (Q^2=10 {\rm~GeV^2})
\right>_{\sigma_L} \simeq 0.25$ fm is similar to that in the
electroproduction of vector mesons $\left< b \left(Q^2=10 {\rm~GeV^2}\right)
\right>_{\gamma^*_L \rightarrow \rho} \simeq 0.35$ fm. However for larger
$Q^2$ the difference between the two values increases and reaches a
factor of 2 for $Q^2 \sim 100$ GeV$^2$.

It can be shown that under certain kinematical conditions the
interaction of a $q\bar q$ pair with the target is given by
equation~(\ref{eq:9c}). In the leading order in $\alpha_s \ln x
\ln{ Q^2\over \Lambda_{QCD}^2}$
the leading Feynman
diagrams for the process under consideration are a hard quark box
diagram with two gluons attached to it and convoluted with the
amplitude for the gluon scattering off a target (see figure~1).

\clearpage
\begin{figure}
\epsfxsize=12cm
\centerline{\epsffile{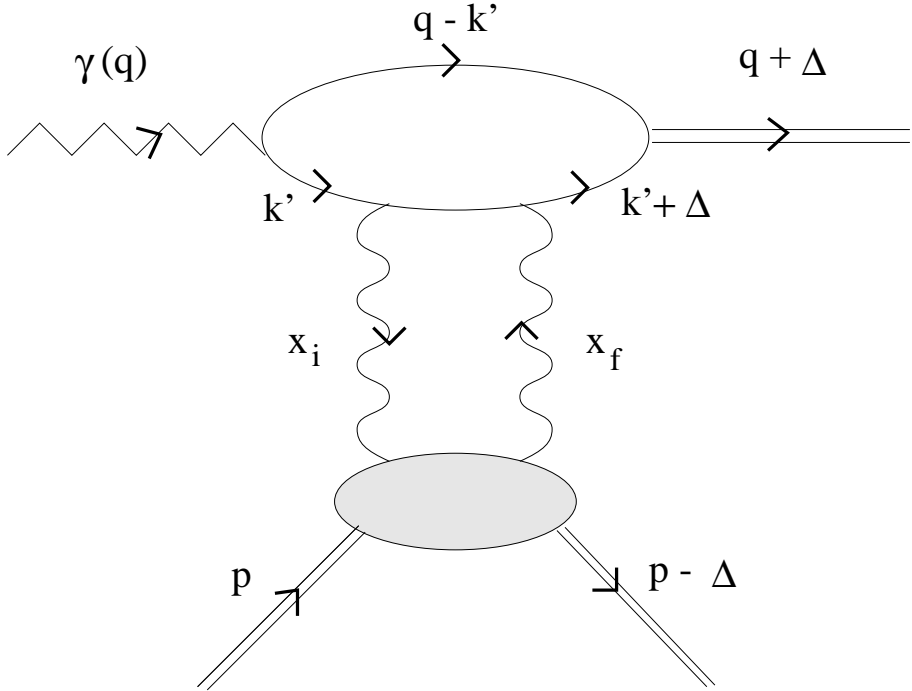}}
\protect{\caption{A typical two-gluon exchange contribution to the
amplitude $\gamma ^*p\rightarrow Vp$.}}
\end{figure}

One can consider the same process in the leading $\alpha_s \ln{
Q^2\over \Lambda_{QCD}^2}$ approximation.  In this case one has to
include also the diagrams where one hard quark line is substituted by
the gluon line. This leads to an extra term $ \propto S_T(x,Q^2)$ in
equation~(\ref{eq:9c}) and allows to treat the parton distributions in
equation~(\ref{eq:9c}) with $\alpha_s \ln{ Q^2\over \Lambda_{QCD}^2}$
accuracy which is more precise than the original leading $\alpha_s
\ln x
\ln{ Q^2\over \Lambda_{QCD}^2}$ approximation in equation~(\ref{eq:9c}).

Since Feynman diagrams are Lorentz invariant it is possible to
calculate the box part of the diagram in terms of the light-cone wave
functions of the vector meson and the photon and to calculate the
bottom part of the diagram in terms of the parton wave function of the
proton.  This mixed representation is different from the QCD improved
parton model which only uses the light-cone wave function of the
target.

The next
step
is to express
this amplitude
through the parton
 distributions
 in the target.
 In the c.m. reference frame of the $ep $ system the target proton has
large momentum $P$, while the photon four-momentum is $\left({\nu-Q^2
\over 4 P},{-\nu-Q^2 \over 4 P},q_t \right)$. The longitudinal momentum
transferred to the proton in this reference frame, $\delta$, is given
by $${\delta \over P} \approx {M_V^2+Q^2 \over \nu}$$. So
the calculation of the imaginary part of the Feynman diagram of figure~1
shows that the fractions of the target momentum
carried by the exchanged gluons $x_i$ and $x_f$ are not equal,
\begin{equation}
x_i-x_f=x,~~~ for~~~ M_V^2 \ll 1.
\label{eq:12c}
\end{equation}
We
neglect terms ${\cal O} ({l_t^2\over Q^2})$ as compared to~1,
with $l_t$ the transverse momentum of the exchanged gluons.
Within the QCD leading logarithmic approximation
\begin{equation}
\alpha_s \ln{1\over x} \sim 1
 \label{eq:13c}
\end{equation}
when terms $\sim \alpha_s$ are neglected, the difference between $x_i$
and $x_f$ can be neglected and the amplitude of the $q \bar q$ interaction
with a target is given by
equation~(\ref{eq:9c})~\cite{BBFS93,FMS93,Brod94}.

We are now able to calculate the cross section for the production of
longitudinally polarized vector meson states when the momentum
transferred to the target $t$ tends to zero~\cite {Brod94},
but $Q^2 \rightarrow \infty $ \footnote{In the paper of Brodsky et at
 \cite{Brod94} the factor 4 in eq.(\ref{eq:14c}) has been missed. We
 are indebted to Z.~Chan and A.~Muller for pointing this out.}

\begin{equation}
\left. {d\sigma^L_{\gamma^*N\rightarrow VN}\over dt}\right|_{t=0} =
{12\pi^2\Gamma_{V \rightarrow e^{+}e^-} m_{V}\alpha_s^2(Q)\eta^2_V~I_V(Q^2)^2|
(xG_T(x,Q^2) + i{\pi\over2}
{d \over d\ln x} xG_T(x,Q)|^2 \over \alpha_{EM}Q^6N_c^2} \ .
\label{eq:14c}
\end{equation}
$\Gamma_{V \rightarrow e^{+}e^-}$ is the decay width of the vector
meson into $e^+e^-$.
The parameter $\eta_V$ is defined as
\begin{equation}
\eta_V\equiv {1\over 2}{\int{dz\over z(1-z)} \Phi_V(z)\over
\int dz \Phi_V(z)}  \ ,
\label{eq:15c}
\end{equation}
where $\Phi_V$ is the light cone wave function of the vector meson.
At large $Q^2$ equation~(\ref{eq:14c}) predicts a $Q^2$ dependence of
the cross section which is substantially slower than $1/Q^6$ because
the gluon densities at small $x$ fastly increase with $Q^2$.
Numerically, the factor $\alpha_s^2(Q^2) G^2(x,Q^2)$
in equation~(\ref{eq:14c}) is
$\propto Q^n$ with $n \sim 1  $ (see figure 4).
An additional $Q^2$ dependence of the cross section arises from the
transverse momentum overlapping intergral between the light-cone wave
function of the $\gamma^*_L$ and that of the vector meson~\cite{FKS},
expressed through the ratio $I_V(Q^2)$
\begin{equation}
I_V(Q^2)= {\int_0^1 {dz \over z(1-z)}\int_0^{Q^2}d^2k_t
{Q^4 \over \left[ Q^2 +{k_t^2+m^2 \over z(1-z)}\right]^2} \psi_V(z,k_t)
\over  \int_0^1 {dz \over z(1-z)}\int_0^{Q^2}d^2k_t \psi_V(z,k_t)}.
\end{equation}
In ref.~\cite{Brod94} it was assumed that $I_V(Q^2)=1$ as for $Q^2
 \rightarrow \infty $ the ratio $I_V(Q^2)$ tends to 1.  But for
 moderate $Q^2$ this factor is significantly smaller than 1.  For
 illustration we estimated $I_V(Q^2)$ for the following vector meson
 wave function: $\psi_V^{(1)}(z,k_t^2)={ c z(1-z) \over
 (k_t^2+\mu^2)^2}$.  The momentum dependence of this wave function
 corresponds to a soft dependence on the impact parameter $b$ - $exp(-\mu
 b)$ in coordinate space.  We choose the parameter $\mu$ so that
 $\left<k^2_t\right> ^{1/2} \in 0.3 \div 0.6~GeV/c$.

Our numerical studies show that the inclusion of the quark transverse
momenta leads to several effects:
\begin{itemize}
\item Different $k_{T}$ dependence of $\psi_V$ leads to somewhat different
$Q^2$ dependence of
$I_V(Q^2)$. Thus electroproduction of vector mesons may become an
effective way of probing $k_t$-dependence of the light-cone $q \bar q$
wave function of vector mesons.

\item The $Q^2$ dependence of $I_V$ for production of vector
mesons build of light quarks $u,d,s$ should be very similar.

\item For  electroproduction  of charmonium states
where $\mu_c \sim \mu {m_{J\Psi }\over m_{\rho}}$ the asymptotic
formula should be only valid for extremely large $Q^2$.
\end{itemize}
The NMC data~\cite{NMC1} and the  HERA data~\cite {ZEUSb} on
diffractive electroproduction of $\rho$ mesons are consistent with
several predictions of equation~(\ref{eq:14c}):
\begin{itemize}
\item a fast increase
with energy of the cross section for electroproduction of vector
mesons as seen in figure~2 from ref.\cite{FKS}
(proportional to $x^{-0.8}$ for $Q^2
=10$~GeV$^2$)~\footnote{This fast increase with decreasing $x$ is
  absent in the non--perturbative two--gluon exchange model of
  Donnachie and Landshoff~\cite{DL1} which leads to a cross section
  rising as $\sim x^{-0.14}$ at $t=0$ and to a much weaker increase of
  the cross section integrated over t.};
\item the dominance of the
longitudinal polarization ${\sigma_L \over \sigma_T}\propto Q^2$;
\item
 the absolute magnitude of the cross section within the uncertainties
 of the gluon densities and of the $k_t$ dependence of the wave
 functions (figure 2)
\item the $Q^2$ dependence of the cross section for
$Q^2 \sim 10 {\rm~GeV^2}$ which can be parameterized as $Q^{-n}$ with $n
\sim 4$.  The difference of $n$ from the asymptotic value of 6 is due
to the $Q^2$ dependence of $\alpha_s^2(Q^2)G_N^2(x,Q^2)$ and of
$I_V^2$ which are equally important in this $Q^2$ range.
\end{itemize}

We discussed above (see also section 8) that the perturbative regime
should dominate in the production of transversely polarized vector
mesons as well, though at higher $Q^2$. This may manifested itself in
the $x$-dependence of the ratio ${\sigma_L \over \sigma_T}$ for fixed
$Q^2$.  At intermediate $Q^2 \sim 10 {\rm~GeV^2}$ where hard physics
already dominates in $\sigma_L$, $\sigma_T$ may still be dominated by
soft nonperturbative contributions.  For these $Q^2$ the ratio should
increase with decreasing $x$ $\sim x^2G_N^2(x,Q^2)$.  At sufficiently
large $Q^2$ where hard physics dominates for both $\sigma_L$ and
$\sigma_T$ the ratio would not depend on $x$.

\begin{figure}
\epsfxsize=15cm
\centerline{\epsffile{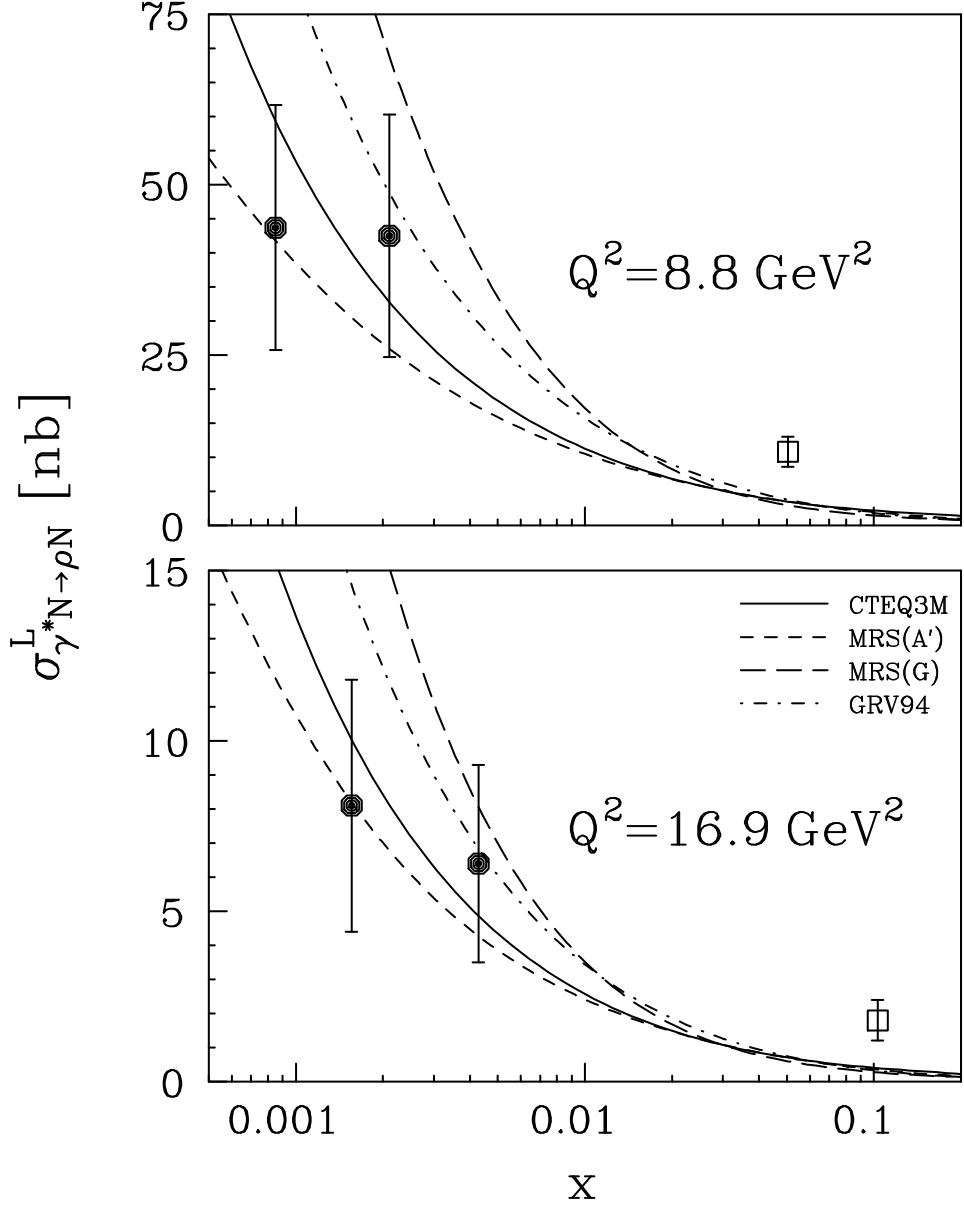}}
\vspace{-1.5cm}
\protect{\caption{The total longitudinal cross section, $\sigma_{\gamma^* N
\rightarrow
\rho N}^L$, calculated from Eq. (\protect\ref{eq:14c}) for several recent
parameterizations of the gluon density in comparison with experimental data
from ZEUS \protect\cite{ZEUSb} (full circles) and NMC \protect\cite{NMC1}
(squares). Typical parameters
for the $\rho$-meson wave functions as discussed above are taken
($\left<k_t^2\right>^{1/2} =0.45 GeV/c$).  We set
$\eta_V=3$ and parameterize the dependence of the differential cross section
on the momentum transfer in exponential form with $B \approx 5$ GeV$^{-2}$.
Note that a change of $T^2(Q^2)$ in the range
corresponding to $\left<k_t^2\right>^{1/2}$ between
0.3 GeV/c and 0.6 GeV/c introduces
an extra scale uncertainty of $0.7 \div 1.4$.
}}
\end{figure}


The t dependence of the cross section is given by the square of the
two gluon form factor of the nucleon $G_{2g}(t)$.  Practically no t
dependence should be present in the block of $\gamma^*$ gluon
interaction for $-t\ll Q^2$.  Thus {\bf the t dependence should be
  universal for all hard diffractive processes}.  Experimentally the
data on diffractive production of $\rho$ mesons for $Q^2 \ge
5{\rm~GeV}^2$~\cite{NMC1}, on photoproduction of $J/\Psi$
mesons~\cite{psi1} and even on neutrinoproduction of $D^*_s$
mesons~\cite{Asratian} show a universal $t$ behavior corresponding to
\begin{equation}
 G_{2g}^2(t) =\exp (Bt)\ \ {\rm with}\ \  B \approx 4{\rm~GeV}^{-2} \ .
\label{eq:15cc}
\end{equation}
A certain weak increase of $B$ is expected with increasing incident
energy due to the so called Gribov diffusion~\cite{Gribovdif}, but
this effect is expected to be much smaller than for soft processes.
However in the limit $Q^2=const$ and $s \rightarrow \infty$ it is
natural to expect an onset of a soft regime, which is characterized
both by a slowing down of the increase of the cross section with
increasing $s$ and by a faster increase of the slope $B$ with $s$,
\begin{equation}
 {\partial \ln B \over \partial \ln s}_ {| s \rightarrow \infty,
Q^2=const} \approx \alpha^{\prime}_{soft} \approx 0.25 GeV^{-2}.
\label{slope1}
\end{equation}
For further discussion see section 15.

We want to point out that for $M_X^2\ll Q^2 $,
the effect of
QCD radiation  is
small.
 This is because bremsstrahlung corrections
due to radiation of hard quarks and gluons
are controlled by the parameter $\alpha_s \ln{x_i\over x_f}$ which is
small since in the reaction considered here $x_i \sim x_f$.  This
argument can be put on a formal ground
within the double logarithmic approximation when only terms $\sim
\alpha_s \ln {1 \over x} \ln {Q^2\over \lambda^2}$ are taken into
account.
One can consider a
more traditional approximation where terms $\simeq \alpha_s
\ln{Q^2\over\Lambda_{QCD}^2}$ are taken into account but terms $\simeq
\alpha_s$ are neglected.

Within these approximations it is legitimate to neglect the
contribution of the longitudinal momentum as compared to the
transverse one. This is a special property of small $x$ physics. Thus
the difference between $x_i$ and $x_f$ leads to an insignificant
correction.

Formula~(\ref{eq:14c}) correctly accounts for nonperturbative
physics and for the diffusion to large transverse distances characteristic
for Feynman diagrams, because  in contrast to the naive applications of the
BFKL Pomeron the diffusion of small size
configurations to large transverse size is not neglected.

  Electroproduction of $J/\Psi$ mesons has been considered also
  in~\cite{Ryskin}
within the nonrelativistic
constituent quark model for $J/\Psi$-meson wave function
in the whole $Q^2$ range. In the limit
  where we can justify the application of PQCD (eq.~(\ref{eq:14c}))
  ($m^2_{J/\Psi}\ll Q^2 $) the result of ref.~\cite{Ryskin}
coincides with
the nonrelativistic limit of our result if $I_V$ is assumed to be equal 1. At
  the same time
the inclusion of the transverse momentum distribution of $c$ quarks in
 the $J/\Psi$ wave function significantly suppresses the cross section
 of the diffractive electroproduction of $J/\Psi$ mesons for $Q^2 \le
 m^2_{J/\Psi}$.  In particular, in the case of photoproduction
 calculations in the model of ref.\cite{Ryskin} which take into account
 the Fermi motion of quarks using realistic charmonium models lead to
 a cross section smaller than the original result by a factor $4 \div
 8$ depending on the model (see discussion in ref.\cite{FKS}).
 Remember that transverse distances essential in the photoproduction
 of the $J/\Psi$ meson are $\sim {3\over m_c}$ which is comparable to the
 average interquark distance in the $J/\Psi$ wave function.
 Since the energy dependence of diffractive photoproduction of
 $J/\Psi$ is consistent with pQCD prediction of \cite{Ryskin}
 the disagreement with the absolute prediction may indicate an important
 role for the interaction with interquark potential.

It is worthwhile to notice the possibility of investigating the
distribution of color in vector meson production in $\gamma^* \gamma$
collisions in reactions like $\gamma^* \gamma \rightarrow V_1 + V_2$
where $V_1$ is produced along $\gamma^*$ and $V_2$ along the
quasi--real photon. Study of this reaction and use of
equation~(\ref{eq:14c}) would allow to measure the gluon density at
small $x$ in various vector mesons. It would be revealing to see how
large is the difference between say $G_{\rho}(x,Q^2)$ and
$G_{\phi}(x,Q^2)$, $G_{J/\Psi}(x,Q^2)$ and to investigate its
dependence on the difference in radii of these vector mesons.

Another interesting process which can be calculated using the
technique discussed above is the production of vector mesons in the
process $\gamma_L^* +p \rightarrow V +X$ in the triple Reggeon limit
when $-t \ge$~few~GeV$^2$ and $ -t \ll Q^2$.  In this kinematical
domain the dominant contribution is due to the scattering of the two
gluons off a parton of the target $g+g+{\rm parton} \rightarrow {\rm
  parton}$.  To avoid the uncertainties related to the vector meson
wave function it is convenient to normalize the cross section of this
process to that of the exclusive vector meson production at $t=0$
\cite{F94}
\begin{eqnarray}
 {{d\sigma^{\gamma_L^* +p \rightarrow V +X}\over dt}\over
\left.{d \sigma^{\gamma_L^* +p \rightarrow V +p}\over dt}
\right|_{t=0}} &
= & {9\over 8 \pi} \alpha_S^2 \left| \ln{Q^2\over k^2} \right|^2  \nonumber \\
&   & {\int\limits^1_y
\left[ G_p(y',k^2)
+{32 \over 81}S_p(y',k^2) \right] dy' \over \left[ xG_p(x,Q^2) \right] ^2}\ ,
\label{larget}
\end{eqnarray}
where $S_p$ is the density of charged partons in the proton, $\nu=2m_N
q_o$, $x=Q^2/\nu$, $k^2=-t$, $y=-t/{2(q_o-p_{Vo})}m_N$ with $p_{Vo}$
the energy of the vector meson and all variables are defined in the
nucleon rest frame.

It follows from equation~(\ref{larget}) that the cross section of the
process $\gamma_L^* +p \rightarrow V +X$ should decrease very weakly
with $t$ and therefore it is expected to be relatively large at
$-t\sim {\rm few~ GeV}^2$.  Similarly to the approach taken
in~\cite{FS89,MuTan} one can easily improve equation~\ref{larget} to
account for leading $\alpha_s \ln x$ terms.

Equation~(\ref{larget}) is a particular case of the suggestion (and of
the formulae) presented in reference~\cite{FS89}, that semi--exclusive
large t diffractive dissociation of a projectile accompanied by target
fragmentation can be expressed through the parton distributions of the
target. The advantage of the process considered here as compared to
the general case is the possibility to prove the dominance of hard
PQCD physics for a longitudinally polarized photon as projectile and
the lack of t dependence in the vertex $\gamma^* +g\rightarrow g + V$.
These advantages allow to calculate the cross section without free
parameters.

Production of transversely polarized vector mesons by real or virtual
photons in the double diffractive process $\gamma_T +p \rightarrow V
+X$ has been calculated recently within the approximation of the BFKL
Pomeron in~\cite{FR}.  The calculation was performed in the triple
Reggeon limit for large t but $s\gg -t$.  Contrary to reactions
initiated by longitudinally polarized photons this calculation is
model dependent; the end point nonperturbative contribution to the
vertex $\gamma_T^* +g \rightarrow g+V$, and therefore to the whole
amplitude, leads to a contribution which is not under theoretical
control.  This problem is familiar to the theoretical discussions of
high $Q^2$ behavior of electromagnetic form factors of hadrons.

\section{Properties of the electroweak production of vector mesons.}

Longitudinal vector meson production is dominated by small inter-quark
distances in the vector meson wave function. Therefore the
factorization theorem can be used to calculate the cross section for
hard diffractive processes in QCD without model assumptions.  For $
M_V^2 \ll Q^2$ all dependence on the quark masses and thus on flavor
is only contained in the light cone wave functions of vector mesons
and not in the scattering amplitudes. This prediction is non trivial
since experimentally the coherent photoproduction of mesons containing
strange or charm quarks is strongly suppressed as compared to the
SU(4) prediction for the ratio of the production cross section for
various vector mesons, which is
\begin{equation}
\rho^o : \omega : \phi : J/\Psi =9 : 1 : 2 : 8 \ .
\label{eq:17c}
\end{equation}
Experimentally the suppression factor is $\approx 4$ for $\phi$-meson
and $\approx 25 $ for $J/\Psi$.  Thus QCD predicts a dramatic increase
of the $\phi/\rho^o $ and $J/\Psi/\rho^o$ ratios at large $Q^2$.

Moreover, the experience with constituent quark models suggests an
additional enhancement of heavier flavor production since for the
heavy quarkonium states the probability for $q$ and $\bar q$ to come
close together is larger.  In fact equation~(\ref{eq:14c}) derived in
QCD predicts for the ratio of production of mesons $V_1$ and $V_2$ at
large $Q^2$ that

\begin{equation}
 \left. {\sigma (\gamma ^*_L + T \rightarrow V_1 +T) \over
\sigma (\gamma ^*_L + T \rightarrow V_2 +T)} \right| _{~Q^2
\gg M_{V_1}^2,\,M_{V_2}^2} =
{M_{V_1} \Gamma_{V_1 \rightarrow e^+e^-}\eta_{V_1}^2(Q^2)\over
M_{V_2} \Gamma_{V_2 \rightarrow e^+e^-}\eta_{V_2}^2(Q^2)} \ .
\label{eq:18c}
\end{equation}
Based on the measured values of $\Gamma_{V \rightarrow e^+e^-}$ and
estimates of $\eta_{V}$ from QCD sum rules~\cite{CZ} we observe that
equation~(\ref{eq:18c}) predicts a significant enhancement of the
$\phi$ and $J/\Psi$ production
\begin{equation}
\rho^o : \omega : \phi
: J/\Psi =9 : 1 : (2*1.0) : (8*1.5)
\label{eq:19c}
\end{equation}
as compared to the SU(4) prediction.
This prediction is valid for $Q^2 \gg m_V^2$ only. Pre--asymptotic
effects are important in the large $Q^2$ range. They significantly
suppress the cross section for production of charmonium states (see
above discussion). Thus the value of the $J/\Psi /\rho$ ratio would be
significantly below the value given by eq.(\ref{eq:19c}) up to very
large $Q^2$. For example the suppression factor is $\sim 1/2$ for $Q^2
\sim 100 GeV^2$ \cite{FKS}.
At the same time it is likely to change very little the predictions for $\rho,
\omega, \phi $-meson production,  since the masses of these
hadrons are quite close and their $q\bar q$ components should be very
similar.

At very large $Q^2$ the $q\bar q
$ wave functions of all mesons converge to a universal asymptotic wave
function with $\eta_V=3$. In this limit further enhancement of the
heavy resonance production is expected
\begin{equation}
\rho^o : \omega : \phi : J/\Psi =9 : 1  : (2*1.2) : (8*3.4) \ .
\label{eq:20c}
\end{equation}
It is important to investigate these ratios separately for the
production of longitudinally polarized vector mesons where hard
physics dominates and for transversely polarized vector mesons where
the interplay of soft and hard physics is more important.

Equation~(\ref{eq:14c}) is applicable also for the production of
excited vector meson states with masses $m_V$ satisfying the condition
that $ m_V^2\ll Q^2 $.  In this limit it predicts comparable
production of excited and ground states.  There are no estimates of
$\eta_V$ for these states but it is generally believed that for
$\rho'$, $\omega'$ and $\phi'$ it is close to the asymptotic value,
and as a rough estimate, we will assume that $\eta_V=\eta_{V'}$. Using
the information on the decay widths from the Review of Particle
Properties~\cite{RPP} we find that
\begin{eqnarray}
\rho(1450):\rho^o \approx 0.3  \nonumber \\
\omega(1420):\omega  \approx 0.3 \nonumber \\
   \rho(1700):\rho^o \approx 1.0\nonumber \\
 \omega(1600):\omega  \approx  1.0 \nonumber \\
\phi(1680) :   \phi \approx 0.6    \nonumber \\
 \Psi' :J/\Psi  \approx  0.5 \ .
\label{eq:21c}
\end{eqnarray}
In view of substantial uncertainties in the experimental widths of
most of the excited states and substantial uncertainties in the values
of $\eta_{V'}$ and the ratio ${I_{V'} \over I_V}$
these numbers can be considered as good to about a
factor of 2.
 The case of $ \Psi'$ where $\Gamma_V $ is well known is less
ambiguous. In this case estimates using charmonium models indicate a
significant suppression as compared to the asymptotic estimate up to
$Q^2 \sim 20 GeV^2$ where this suppression is $\sim 0.5$ \cite{FKS}.

In spite of these uncertainties it is clear that
 a substantial production of excited resonance states is
expected at large $Q^2$ at HERA. A measurement of these reactions may
help to understand better the dynamics of the diffractive production
as well as the light-cone minimal Fock state wave functions of the
excited states. It would allow also to look for the second missing
excited $\phi $ state which is likely to have a mass of about 1900 MeV
to follow the pattern of the $\rho,~\omega,~J/\Psi$ families.

The relative yield of the excited states induced by virtual photons is
expected to be higher than for real photons since the Vector Dominance
Model (VDM) and equation~(\ref{eq:14c}) lead to
\begin{eqnarray}
\left. {\sigma (\gamma  + N \rightarrow V +N) \over
\sigma (\gamma   + N \rightarrow V' +N)}
 {\sigma (\gamma ^*_L + N \rightarrow V' +N) \over
\sigma (\gamma ^*_L + N \rightarrow V  +N)}
\right| _{~Q^2 \gg M_{V'}^2,M_{V}^2}
=\nonumber \\
={M_{V'}^2 \over M_V ^2}  {  \eta_{V'}^2(Q^2)\over \eta_{V}^2(Q^2)}
{\sigma^2_{tot}(V'N)
\over \sigma^2_{tot}(VN)}   \ge {M_{V'}^2 \over M_V ^2}  \ .
\label{eq:22c}
\end{eqnarray}
In the last step we used an empirical observation that for effective
cross sections of $V'N$ and $VN$ interactions which enter in the VDM
model~\footnote{Note that these effective cross sections have no
  direct relation to the genuine interaction cross sections.  For
  example, based on geometrical scaling one expects the interaction
  cross section to increase with the size of the projectile
  approximately as $R=\sigma_{tot} (\Psi'N)/\sigma_{tot}(J/\Psi N)
  \sim R^2_{\Psi'}/R^2_{J/\Psi} \sim 4$.  However if one applies the
  equations of the VDM for the extraction of the cross sections from
  photoproduction of $J/\Psi,~\Psi'$ one finds $R \sim 0.7 \sim
  M^2_{J/\Psi}/M^2_{\Psi'}$.  This trend seems to reflect effects of
  color screening in the production of heavy quarkonium
  states~\cite{FS85,FS88}.  Note also that photoproduction data do not
  resolve $\rho(1430)$ and $\rho(1700)$.  In the case of $\rho'$
  photoproduction off nuclei similar nuclear absorption effects are
  observed for the production of $\rho$ and $\rho'$, indicating
  $\sigma(\rho'N) \approx \sigma(\rho N)$.  At the same time
  application of the vector dominance model for the process $\gamma~p
  \rightarrow \rho' p$ leads to $\sigma(\rho'N) \approx 0.4
  \sigma(\rho N)$.  The observed pattern indicates that production of
  $\rho'$ is dominated by average quark-gluon configurations (large
  absorption cross section), while the probability of these
  transitions is suppressed since the transition $\gamma^*\rightarrow
  V$ emphasizes the role of small configurations.} ${\sigma_{tot}(V'N)
  \over \sigma_{tot}(VN)} \leq$1and that $\eta_V$ and $\eta_{V'}$ are
close to the asymptotic values for light mesons while for heavy quark
systems the values of $\eta_V'$ are close to the static quark value of
$\eta_V=2$.

Another interesting QCD effect is that the ratio of the cross section
for the diffractive production of excited and ground states of vector
mesons should increase with decreasing $x$ and $Q^2$. This is because
the energy denominator - ${1 \over \left({m_q^2+k^2_t \over z(1-z)} -
m_{V'}^2\right)}$, relevant for the transition $V \rightarrow q \bar
q$ (with no additional partons) should be large and positive. Thus
the heavier the excited state, the larger Fermi momenta should be
important. Thus the gluon distributions should enter at larger
virtualities in the case of $V'$ production.

Equation (\ref{eq:14c}) is applicable also to vector meson
production in weak processes.  Consider for example the diffractive
production of $D^{*\pm}_s=c\bar s$ meson in $ W^{\pm}N$ scattering. To
calculate this cross section it is sufficient to substitute in
equation~(\ref{eq:14c}) the electromagnetic coupling constant by
$g \cos\theta_C$, where $\theta_C$ is the Cabibbo angle.  Some
enhancement of the $D^*_s$ production is expected due to a larger
value of $\eta_{D^*_s}$ originating from the asymmetry in the $x$
distribution of the light and heavy quark in $D^*_s$.

To summarize, the investigation of exclusive diffractive processes
appears as the most effective method to measure the minimal Fock
$q\bar q$ component of the wave functions of vector mesons and the
light-cone wave functions of any small mass hadron system having
angular momentum 1. This would be very helpful in expanding methods of
lattice QCD into the domain of high energy processes.

\section{Color transparency phenomenon.}

\subsection{Coherent production of vector mesons off nuclei  at small t}

The QCD analysis described in section 4 confirms the conjecture of
refs.~\cite{FS88,BM88} that at large $Q^2$ vector mesons are produced
in small transverse size configurations (at least for the
longitudinally polarized photons) and hence the color transparency
phenomenon (CT) is expected.  In the case of coherent vector meson
production off nuclear targets QCD prediction, in the form of
equation~(\ref{eq:14c}), absorbs all the dependence on the atomic
number in the gluon
and sea quark distributions
 of the target. But it is well known
that the evolution of parton distributions with $Q^2$ moves the effect
of nuclear shadowing to smaller $x$.  Thus at small but fixed $x$ and
sufficiently large $Q^2$ the cross section for hard diffractive
processes is expected to fulfill the following relation:
\begin{equation}
\left. {d\sigma^L_{\gamma^*+A \rightarrow X+A}\over dt}\right| _{t=0} =
A^2 \left. {d\sigma^L_{\gamma^*+N \rightarrow X+N}\over dt}\right| _{t=0} \ .
\label{eq:18cc}
\end{equation}
This is the so called color transparency phenomenon which leads to the
validity of the impulse approximation -- the nucleus is transparent
for the projectile and there is no final state interactions. The onset
of CT should occur at moderate $Q^2$ since gluon shadowing disappears
  fast  with increasing $Q^2$ at fixed $x$~\cite{FLS}. If the size of
the configuration is fixed (at large but fixed $Q^2$) but the energy
of the collision increases, shadowing effects should become more and
more important since the gluon shadowing increases with decreasing
$x$~\cite{FS88}, (see figure~3).
Moreover the analysis of the unitarity
constraints in section 11 demonstrates that
in the scattering off heavy nuclei screening effects should lead to
very substantial suppression of coherent vector meson
production cross section
$\left. {d\sigma^L_{\gamma^*+A \rightarrow X+A}\over dt}\right|_{t=0} $
for $x \sim 10^{-4}, Q^2 \sim {\rm~10 GeV^2}$
as compared to the expectation of eq.(\ref{eq:18cc}).
  We shall explain in
the next section that a similar CT behavior is expected for the
production of transversely polarized vector mesons but at
significantly larger $Q^2$ than for the longitudinally polarized
vector mesons.

\begin{figure}
\centerline{\epsffile{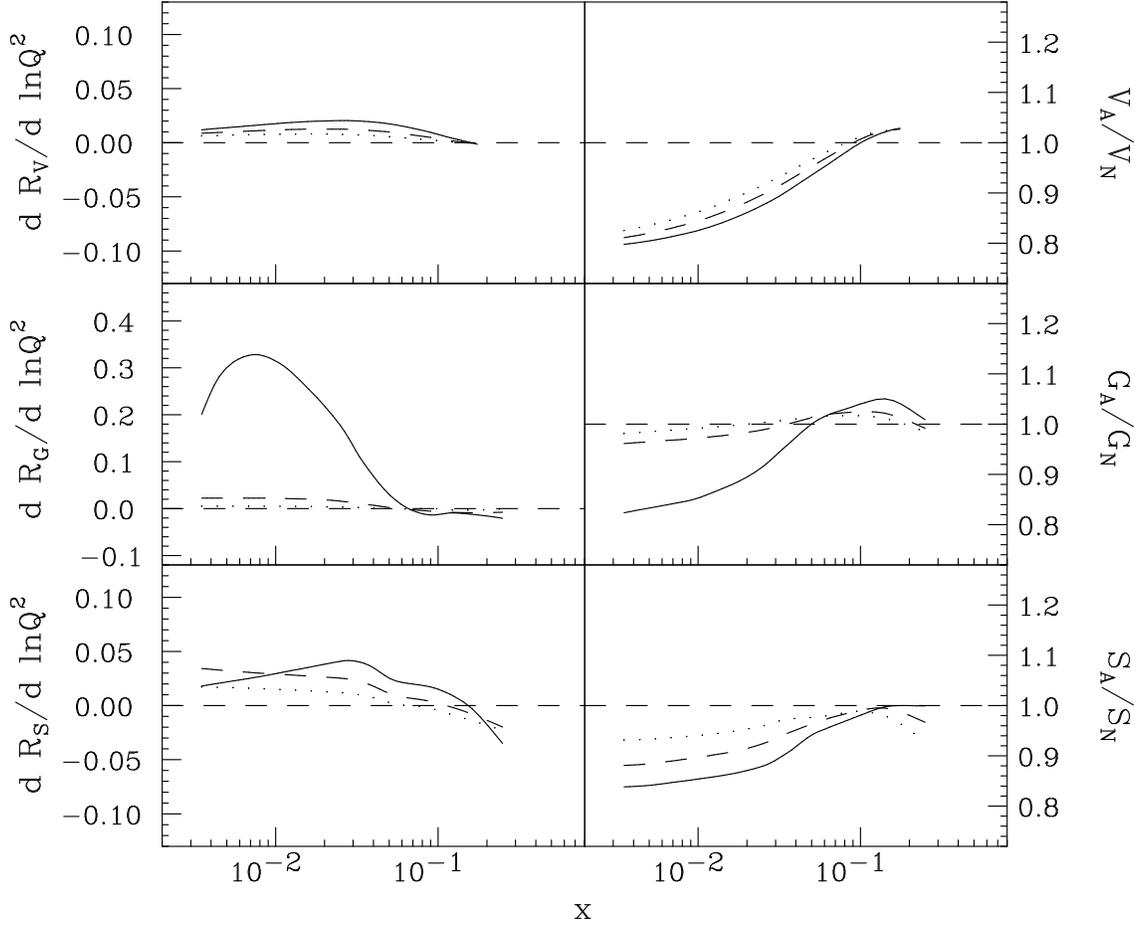}}
\protect{\caption{The ratios $R_S$, $R_V$, and $R_G$ of sea, valence
    and gluon distributions for A=40 and A=2 and (left) their
    logarithmic derivatives, $d R_{S(V,G)}(x,Q^2) /d (\ln Q^2)$
    as a function of $x$ for $Q^2 = 4$~GeV$^2$ (full line), $Q^2 = 25
    {\rm~GeV}^2$ (dashed line) and $Q^2 = 100 {\rm~GeV}^2$ (dotted line).}}
\end{figure}

In contrast to the formula derived in QCD (equation~(\ref{eq:14c}))
the two gluon exchange constituent
quark model~\cite{GS} predicts no increase of
the cross section at small $x$. Quark models and the Glauber
approximation (used in particular in~\cite{KNNZ93}) are in variance
with
the factorization theorem in QCD
even if formula~(\ref{eq:9c}) obtained in
{}~\cite{BBFS93,FMS93} is used for the nucleon target since they predict
the disappearance of shadowing at fixed $Q^2$ when $x$ decreases.

As explained above, the preliminary HERA data indicate that PQCD
predictions contained in formula~(\ref{eq:14c}) are applicable already
for $Q^2 \sim 10$~GeV$^2$.  Obviously this is an implicit
confirmation of the color transparency logic since it confirms both
the presence of small transverse configurations in the $\rho$ meson
and the smallness of their interactions with hadrons.  It would be
important to investigate further these effects more directly at ultra
high energies. To this end we consider briefly the scattering off the
lightest nuclei \cite{FMS95}.  Note that there are discussions to
accelerate deuterons at HERA and to polarize them in order to measure
the parton distributions in the neutron.

\subsection{Color transparency effects in  $\gamma^*_L
 + D (A) \rightarrow V_L +D(A)$.}


The very existence of the color coherence effects leads to
a rather nontrivial dependence of the
cross sections of hard diffractive processes on $x,Q^2$. To
elucidate this point we consider in this section  diffractive
 electroproduction of vector mesons off the deuteron.

First, let us consider the leading twist effect. It follows from
eq.(\ref{eq:18cc}) that at t=0 the amplitude of this process is proportional to
the parton density in the deuteron. The nuclear effect in the leasing
twist depend on $x,Q^2$ in a rather complicated way. At $x \sim 0.1$
and $Q^2 \sim ~~ few~~ GeV^2$ - the kinematics
of the HERMES facility the gluon density in nuclei is significantly
enhanced:$ {G_A(x,Q^2) \over A G_N(x,Q^2)} > 1$. This effect follows
from the need to reconcile the momentum and baryon sum rules with the
${F_{2A}(x,Q^2) \over F_{2N}(x,Q^2)}$ data \cite{FLS}.
The dynamical mechanism relevant for the gluon
enhancement is not understood so far.
One example of the Feynman diagrams which may lead
 to such an enhancement is the
propagation of color states in nuclei and their mixing with
 multinucleon states via gluon exchanges.
Since the gluon exchange amplitude is real the  contribution
of such diagrams  corresponds to enhancement of the gluon field not the
shadowing. Legitimate calculations of such a mechanism are absent at
present.

Consequently QCD predicts an enhancement but no shadowing for
the electroproduction of vector mesons at $t=0$ off the deuteron
at $x \sim 0.1$.
This effect
should die out rather rapidly with increase of $Q^2$ due to the QCD evolution
of parton distributions with $Q^2$ (cf. Fig.  for the $Q^2$ dependence
of parton distributions).

At sufficiently small $ x \le 10^{-2}$ shadowing of gluon distribution
dominates.
We will restrict the discussion to the region of sufficiently large
$x \ge 10^{-4}$ where   interaction of a small $q \bar q$ state with a nucleon,
 $\sigma_{q \bar qN}(b^2,x)$ which is given by eq.(\ref{eq:9c}) is small as
compared to the unitarity limit and therefore
 QCD evolution equations seem to be
applicable. In this kinematics one expects a fast decrease
of shadowing with increase of $Q^2$.

Obviously, at $t \approx t_{min}$ shadowing effects are
small since inter-nucleon distance in the deuteron are comparatively
large. To enhance these effects it would be advantageous to study
experimentally the coherent electroproduction of
vector mesons at $|t| \ge 0.5 GeV^2$ where an interesting
diffraction pattern with secondary maximum
 was observed long time ago for photoproduction of $\rho$-meson.
This pattern at $Q^2=0$ arises within the vector dominance model
as a result of the vector meson rescatterings. At large $Q^2$ QCD predicts more
complicated behavior.

Let us consider first the rescatterings of the produced $q \bar q$
pair of small size $b$.
The scattering amplitude is given by the sum of two terms, one given
by the impulse approximation and the other due to double
scattering~\footnote{Since the relevant $b$ are small we neglect here
in the first approximation effects of the leading twist nuclear
shadowing induced by the space-time evolution of the $q \bar q$ pair
leading to the formation of large spatial size quark-gluon
configurations.}
\begin{eqnarray}
 & & {d\sigma_L(\gamma^*+D \rightarrow V+D)\over dt} =
 {1 \over 16 \pi}\left|\int  \left[ 2S_D(t)f_{\gamma^*N \rightarrow
VN}(x,b^2,r_t)
+\nonumber \right. \right.\\
 & & \int {i \over 8 \pi^2} \nonumber \left.
f_{\gamma_L^*N \rightarrow VN}(x,b,Q^2, r_t/2-k_t)
 f_{q \bar q, N}(x,b, r_t/2 + k_t)S_D(4k_t^2)\right.\\ & & \left.\left.
 \psi_{\gamma^*}(z,b,Q^2)
 \psi_V(z,b) dz  d^2k_t \right] d^2b
\right| ^2 \ ,
\label{eq:18c1}
\end{eqnarray}
where $t=-r_t^2$, $S_D(t)$ is the deuteron form factor, and $f_{q \bar
  q, N}(x,b,r_t/2+k_t)$ is the amplitude for the elastic
rescattering of the $q \bar q$ pair.  For simplicity we ignore here the
spin indices.  For the interaction of a small transverse size $q \bar
q$ configuration small impact parameters $b$ dominate in
equation(\ref{eq:18c1}).  Hence the CT prediction of
formula~(\ref{eq:9c}) is that at small but fixed $x$ with increasing
$Q^2$ the relative contribution of the second term should be
proportional to ${1\over Q^2}xG_N(x,Q^2)$.  Since at $-t \ge -t_0 \sim
0.5 {\rm~GeV}^2$ the elastic cross section is dominated by the
square of the second term,
this mechanism leads
in this region to a cross
section which is
extremely sensitive to the CT effects. In particular,
the ratio
\begin{eqnarray}
\left. {d \sigma^{\gamma_L^* + D
\rightarrow V +D}\over dt }\right| _{-t \ge -t_0} /
\left. {d \sigma^
{\gamma_L^* + D   \rightarrow V +D} \over dt }\right| _{-t=0}  =
\left| {\left<\sigma_{q\bar qN}(b)\right>
 \over 4 \pi} \left< {1\over R^2} \right> \right| ^2
{\exp Bt \over 4}
\nonumber \\ \propto {x^2 G^2_N(x,Q^2) \exp Bt \over Q^4},
\label{gammad}
\end{eqnarray}
where
\begin{equation}
\left<\sigma_{q\bar qN}(b)\right> = {\int d^2b
\psi_{\gamma^*_L}(b) \psi_{V}(b)\sigma^2_{q\bar qN}(b)
\over \int d^2b \psi_{\gamma^*_L}(b) \psi_{V}(b)\sigma_{q\bar qN}(b)}
\label{sigm2}
\end{equation}
should strongly decrease with increasing $Q^2$ and flatten for
sufficiently large $Q^2$ to a leading twist behavior due to the
space-time evolution of the $q \bar q$ configurations. On the contrary
at fixed $Q^2$ this ratio should increase with decreasing $x$. Here
$B=B_{\gamma^* N}/2$ with $B_{\gamma^* N}$ denoting the slope of the
differential cross section for the elementary $\gamma^* + N
\rightarrow V + N$ reaction and $\left<{1\over R^2}\right> =\int d^3r
r^{-2}\psi_D^2(r)$.  The large $t$ ($-t \ge 0.5 {\rm~GeV}^2$) dependence
of the cross section in equation~(\ref{gammad}), ${d \sigma \over dt}
\propto \exp(B't)$ with $B' \sim 2 {\rm~GeV}^{-2}$, is significantly
weaker than in the Glauber model where $B'$ is expected
to be
\begin{equation}
B'={B_{\gamma^* N} B_{\gamma N} \over  B_{\gamma^* N} + B_{\gamma N}}
\approx 3 {\rm~GeV}^{-2 } \ .
\label{slope}
\end{equation}
We neglected here the deuteron quadrupole form factor effects. They
lead to a contribution to the cross section which does not interfere
with the electric transition and for which Glauber effects are small.
This contribution fills the minimum due to the interference of the
impulse and double scattering terms~\cite{Franco}. However this
contribution to the cross section can be significantly suppressed by
using a polarized deuteron target. Similar effects should be present
for the scattering off heavier nuclei, like $^{3,4}He$. The
measurement of the depth of the Glauber minimum due to the
interference of the amplitude given by the impulse approximation with
rescattering amplitudes would allow to check another feature of
expression~(\ref{eq:14c}), namely the large value of the real part of
the production amplitude $Re f/Im f \sim \pi n/2 \sim 0.5$, where $n$
characterizes the rate of increase of the gluon density at small $x$,
$ xG_N(x,Q^2) \propto x^{-n}$.
In this discussion we neglected the leading twist mechanism of double
rescattering related to the leading twist nuclear shadowing.  It is
likely to have a similar $t-$dependence as the term considered
previously. It may compete with the mechanism we discussed above in a
certain $x,Q^2$ range. This question requires further studies. In any
case it is clear that in a wide kinematic range the relative height of
the secondary maximum would be strongly suppressed as compared to
the case of vector meson production by real photons.  At very
small $x$ for $Q^2$ where $\sigma_{q \bar q N}$ is close to unitarity
bound this suppression may disappear. This would establish the $x,Q^2$
range where color transparency should disappear.

Recent FNAL data on incoherent diffractive electroproduction of vector
mesons off nuclear targets~\cite{FNAL94} did find an increase of
nuclear transparency with increasing $Q^2$ as predicted
in~\cite{FS88,BM88,KNNZ93}.
However in this experiment the increase of $Q^2$ is correlated with
the increase of the average $x$ and a significant effect is reported for a
$Q^2$ range where $\left< x \right>$ corresponds to
average longitudinal distances which are comparable with the nuclear
radius $l_c={1 \over 2 m_N x} \sim R_A$. It is well known that at
large $x$ shadowing disappears for hard processes. Thus it is
necessary to investigate theoretically to what extent the observed
increase of transparency is explained by the effects of finite
longitudinal distances. The ideas discussed in this report do not
apply directly to color transparency phenomena at moderate energies.
For a recent review of this field we refer the interested reader
to~\cite{FMS94}.

\section{Electroproduction of photons.}

The diffractive process $\gamma^* + p \rightarrow \gamma + p$ offers
another interesting possibility to investigate the interplay between
soft and hard physics and to measure the gluon distribution in the
proton.  We shall consider the forward scattering in which case only
the transverse polarization of the projectile photon contributes to
the cross section.  This follows from helicity conservation.  In this
process, in contrast to reactions initiated by longitudinally
polarized highly virtual photons, soft (nonperturbative) QCD physics
is not suppressed. As a result, theoretical predictions are more
limited. Within QCD one can calculate unambiguously only the
derivative of the amplitude over $\ln{Q^2\over Q_o^2}$ but not the
amplitude itself. However for sufficiently small $x$ and large $Q^2$,
when $\alpha_s(Q_o^2) \ln{Q^2\over Q_o^2} \ln x$ is large, PQCD
predicts the asymptotic behavior of the whole amplitude.

It is convenient to decompose the forward scattering amplitude for the
process $\gamma^* + p \rightarrow \gamma + p$ into invariant structure
functions in a way similar to the case of deep inelastic
electron-nucleon scattering. Introducing the invariant structure
function $H(x,Q^2)$, an analogue of $F_1(x,Q^2)$ familiar from deep
inelastic electron scattering off a proton, we have
\begin{equation}
\left. {d\sigma  \over dt}^{\gamma^*+N \rightarrow \gamma + N}
\right| _{t=0} =
\pi \alpha_{em}^2 {H(x,Q^2)^2\over {s^2}} \ .
\label{eq:26c}
\end{equation}

When $Q^2$ is sufficiently large, QCD allows to calculate the $Q^2$
evolution of the amplitude in terms of the parton distributions in the
target. As in the case of deep inelastic processes it is convenient to
decompose $H(x,Q^2)$ in terms of photon scattering off flavors of type
$i$
\begin{equation}
H(x,Q^2)=\sum\limits_{i}  e_i^2 h_i(x,Q^2) \ ,
\label{eq:27c}
\end{equation}
where the sum runs over the different flavors $i$ with electric charge
$e_i$. It is easy to deduce the differential equation for $h_i$, the
analogue of the evolution equation for the parton distributions.
\begin{eqnarray}
  {dh_i(x,Q^2) \over d\ln{Q^2 }} = {\alpha_{s}(Q^2)\over 2\pi} \int
  {dz\over z} \left[ P_{qg}\left( {x\over z}\right) G_p(z,Q^2) +
  \nonumber \right. \\ \left.  P_{q q}\left( {x\over z}\right) q_i(z,
  Q^2)\right] \left[ 1 + {x\over z}\left( 1-{x\over z}\right) \right]
  +{\cal O}(\alpha_s^2) \ .
\label{eq:28c}
\end{eqnarray}
Here $P_{qq}$ and $P_{qg}$ are the splitting functions of the GLDAP
evolution equation~\cite{GLDAP} . The factor $ 1 + {x\over z}\left(
1-{x\over z} \right)$ takes into account the difference of the
virtualities of the initial and final photon. The solution of this
equation is
\begin{eqnarray}
& & h_i(x,Q^2)  =  h_i(x,Q_{0}^2) +{\alpha_{s}(Q^2)\over 2\pi}
\int\limits_{\ln Q_o^2}^
{\ln Q^2}d\ln Q_1^2 \int\limits_{x}^{1} {dz\over z} \nonumber \\
& &\left[ P_{qg}\left( {x\over z}\right) G_p(z,Q_1^2) +
P_{q
  q}\left( {x\over z} \right) q_i(z, Q_1^2) \right] \left[ 1 + {x\over
  z}\left( 1-{x\over z}\right) \right] +{\cal O} (\alpha_s^2) \ .
\label{eq:29c}
\end{eqnarray}
Usually it is assumed that the soft components of the parton
distributions increase at small $x$ more slowly than the hard ones. If
this is the case, at sufficiently small $x$, in the leading $\alpha_s
\ln x$ approximation, the first term in equation~(\ref{eq:29c}) can be
neglected.  As a result one can obtain the asymptotic formula for the
whole $H(x,Q^2)$ and not only for its derivative.

Similarly to the case of electroproduction of photons it is not
difficult to generalize the $Q^2$ evolution equation to the amplitude
for the diffractive production of transversely polarized vector mesons.
One of the consequences of this evolution equation is that, at
asymptotically large $Q^2$ and small $x$, the production cross section
has the same dependence on the atomic number of a target as in the
case of longitudinally polarized vector mesons.

\section{Coherent Pomeron.}

It is interesting to consider high-energy hard processes in the
diffractive regime with the requirement that there is a large rapidity
gap between the diffractive system containing the high $p_t$ jets and
the target which can remain either in the ground state or convert to a
system of hadrons. In PQCD such a process can be described as an
exchange of a hard gluon accompanied by a system of extra gluons which
together form a color neutral state.  It was predicted~\cite{FS89}
that such processes should occur in leading twist. ( Note that in
reference~\cite{MR} it was stated that this process should rather be a
higher twist effect.  This statement was due to some specific assumptions
about the properties of the triple Pomeron vertex).

The simplest example is in the triple Reggeon limit the production of
high $p_t$ jets in a process like
\begin{equation}
h + p \rightarrow jet_1 + jet_2 +X +p
\label{eq:30c}
\end{equation}
where the final state proton carries practically the whole momentum of
the initial proton.  The initial particle can be any particle
including a virtual photon.  To probe the new PQCD hard physics the
idea~\cite{FS89} is to select a final proton with a large transverse
momentum $k_t$. One can demonstrate that this selection tends to
compress initial and final protons in small configurations at the
moment of collision. In this case the use of the PQCD two gluon
exchange or two--gluon ladder diagrams becomes legitimate. A
nontrivial property of these processes is a strong asymmetry between
the fractions of the target momentum carried by the two gluons (the
contribution of the symmetric configurations is a higher twist effect
with the scale determined by the invariant mass of the produced two
jets~\cite{CFS93}). Thus one expects gluon bremsstrahlung to play a
certain role \cite{F92}.  However since the proton is in a
configuration of a size $\sim {1 \over k_t}$ this radiation is
suppressed by the small coupling constant: $\sim \alpha_s(k_t^2) \ln
({p_t^2\over k_t^2})$.  When $k_t$ tends to 0 this radiation may
suppress significantly the probability of occurrence of events with
large rapidity gaps.

The prediction is that such a process appears as a leading twist
effect \cite{FS89}
\begin{equation}
{d\sigma\over dp_t^2} \sim {1\over p_t^4} \ .
\label{eq:31c}
\end{equation}
This prediction is in an apparent contradiction with a naive
application of the factorization theorem in QCD which states that the
sum of the diagrams with such soft gluon exchanges cancels in the
inclusive cross section. However in reaction~(\ref{eq:30c}) we
selected a certain final state with a white nucleon hence the usual
proof of the factorization theorem does not hold anymore --- there is
no cancelation between absorption and radiation of soft
gluons~\cite{CFS93}.  This conclusion was checked in a simple QED
model with scalar quarks~\cite{SB}.

It was suggested by Ingelman and Schlein~\cite{IS} to consider
scattering off the Pomeron as if the Pomeron were an ordinary particle
and to define parton distributions in the effective Pomeron. In this
language the mechanism of hard interaction in diffraction discussed
above would contribute to the parton distribution in the Pomeron a
term proportional to
  \begin{equation}
 \delta(1-x) \ \ \  {\rm or}  \ \ \ \    {1\over (1-x)} \ \ .
\label{eq:33c}
\end{equation}
This term corresponds to an interaction in which the Pomeron acts as a
whole. Hence the term coherent Pomeron.  In this kinematical
configuration the two jets carry practically all the longitudinal
momentum of the Pomeron.  The extra gluon bremsstrahlung discussed
previously renders the $x$ dependence somewhat less singular at $x
\rightarrow 1$ but the peak should be concentrated at large
$x$~\cite{F92,CFS93}.  There are no other known mechanisms generating
a peak at large $x$.  The recent UA(8) data~\cite{Br} on the reaction
$p +\bar p \rightarrow jet_1 + jet_2 +X +p$, with the proton
transverse momentum in the range $2{\rm~GeV}^2\geq k_t^2\geq
1{\rm~GeV}^2$, seem to indicate that a significant fraction of the two
jet events corresponds to the $x \sim 1$ kinematics.  It is thus
possible that the coherent Pomeron contributes significantly to the
observed cross section~\footnote{ The coherent production of high
  $p_t$ jets by a real photon has been first discussed by Donnachie
  and Landshoff ~\cite{DL2} and then rediscussed in
  reference~\cite{XY}.  This process, discussed in the next section,
  gives a negligible contribution in the kinematic regime
  characteristic for the coherent Pomeron.}.

 The prediction is that the contribution of the coherent
Pomeron to diffractive  electroproduction of dijets
at $p_t^2 \gg Q^2$ should be suppressed by an additional power of $Q^2$
$${d \sigma^{\gamma^* +p \rightarrow 2 jets +X + p} \over dp^2_t} \sim
{1 \over p^4_t} {1 \over Q^2}$$ as compared to
$${d \sigma^{\gamma^* +p \rightarrow 2 jets +X } \over dp^2_t} \sim {1
  \over p^4_t} $$ for other hard processes originating from the hard
structure of the virtual photon.

The complicated nature of the
effective Pomeron should manifested itself in several ways in hard
diffraction~\cite{FS89,CFS93}.\\
(i) There should be a significant suppression of the coherent Pomeron
mechanism at small $t$ due to screening (absorptive) effects since at
small $t$ the nucleon interacts in an average configuration.  This
suppression should be larger for $pp$ scattering than for $\gamma p$
scattering since absorptive corrections increase with the increase of
the total cross section (for $\gamma p$ interaction the VDM effective
total cross section at HERA energies is $\leq 30$ mb).\\
(ii) Due to the contribution of soft physics, the effective Pomeron
structure function as determined from the low $t$ diffractive
processes should be softer than for large $t$ diffraction.

Therefore it would be very important to compare hard diffractive
processes induced by different projectiles and to look for deviations
from the predictions based on the simplest assumption that the Pomeron
has an universal parton distribution~\cite{CTEQ}.

\section{Forward electroproduction of jets.}

Forward diffractive photo and electroproduction of high $p_t$ jets off
a nucleon target (in the photon fragmentation region) $\gamma^* + N
\rightarrow jet_1 +jet_2 +N$ is another promising process to
investigate the interplay of soft and hard physics. We shall confine
our discussion to the kinematical region
\begin{equation}
{-\left<r_N^2\right> t_{min}\over 3}=
\left({Q^2+M_{q\bar q}^2 \over 2q_0}
\right)^2
{\left<r_N^2\right> \over 3} \ll 1,
\label{eq:33c1}
\end{equation}
where
\begin{equation}
 M_{q\bar q}^2 = {(m_q^2 + p_t^2)\over z(1-z)}
\label{eq:34c1}
\end{equation}
is the square of the invariant mass of the  produced $q\bar q$ system, $m_q$
is the mass of quarks and $z$ is the fraction of photon momentum carried
by the $q$ or $\bar q$. In this regime the coherence of the produced hadron
states allows to express the amplitude through the gluon distribution in
the target.

An interesting effect occurs in the photoproduction
 of high $p_t$ jets.
 The contribution of a single Feynman diagram with the 2 gluon
exchange in the $t$ channel contains terms $R_1\approx {p_{t
{}~\mu}\over {p_t^2+M^2}}$ and $R_2\approx { m\over p_t^2}$.  Here $m$
is the mass of a bare quark, $M$ can be calculated through $m$ in pQCD
but in general accounts for the nonperturbative physics. We omit
constants and $\sigma$ matrixes in this dimensional estimate and
restrict ourselves to the contribution of large $p_t$ only.  A
cancelation occurs when the sum of diagrams is considered. It
accounts for the fact that the sum of diagrams describes the
scattering of a colorless dipole.

   Naively we should expect that after cancelation $R_1$ term
should become $R_1\approx {p_{t~\mu}\over (p_t^2+M^2)^2}$. But in reality
it becomes $R_1\approx {M^2 p_{t~\mu}\over (p_t^2+M^2)^3}
\approx {1 \over p_t^5}$.
$R_2$ term after cancelation in the sum of diagrams becomes
$R_2\approx  {m\over {p_t^4}}$.
            Thus cross section of forward photoproduction
of $q \bar q$ pair $d\sigma/dt dp_t^2$ contains terms:
   $m^2\over p_t^8$ \cite{XY},  ${M^4\over p_t^{10}}$
       and ${M^2 m\over p_t^9}$.

  Since mass of light quark is small it is reasonable to put it 0.
It is not legitimate to put $M=0$. So expected asymptotical behavior is
${M^4\over p_t^{10}}$.
 Thus photoproduction of charm
should dominate hard diffractive photoproduction processes for $ p_t
\geq m_c$ \cite{XY}.

Photoproduction of high $p_t$ jets originating from the
fragmentation of light flavors is predominantly due to next to leading
order processes in $\alpha_s$.

The diffractive electroproduction of dijets seems to be the dominant
process in the region of $M_{q\bar q}^2\leq Q^2$, while in the region
$M_{q\bar q}^2\gg Q^2$ exclusive dijet production is one of many
competing processes contributing to the diffractive sector like
radiation of gluons from quark and gluon lines.

In the approximation when only leading $\alpha_s \ln x$ terms are kept,
the  off mass shell effects in the amplitude for the $q\bar q$ interaction
with a target are unimportant. Therefore  the  total cross section  of
diffractive electroproduction of jets by longitudinally polarized
photons can be calculated by applying  the optical theorem for the elastic
$q\bar q$ scattering off a nucleon target and equation~(\ref{eq:9c}) for
the total cross section of $q\bar q$ scattering off a nucleon:
\begin{equation}
\sigma(\gamma_L^* + N \rightarrow jet_1 +jet_2 +N)= {1\over 16\pi B}
\int \psi^2_{\gamma_L^*}(z,b)\cdot(\sigma(b^2))^2 dz d^2b
\label{eq:34d1}
\end{equation}
Here B is the  slope of the two gluon form factor  discussed in section~4
and $\psi_{\gamma_L^*}(z,b)$ is the wave function of the longitudinally
polarized photon.
Essentially the
 same equation is valid for the production by
transversely polarized virtual photons of two jets which share equally the
momentum of the projectile photon.

In~~~\cite{Ryskin2} it has been assumed that diffractive production of
jets off a proton is a hard process at each stage.  The formula obtained
under this assumption resembles equation~(\ref{eq:34d1})
with leading
 $\alpha_s \ln {Q^2\over \Lambda^2_{QCD}} \ln {1 \over x}$
with the PQCD part of the gluon distribution in a target.
In view of the nontrivial interplay of soft and hard physics of large
longitudinal distances this approach
is questionable in QCD
if $x$ is not extremely small.
It is most easily seen when one considers the effect
of nuclear shadowing in diffractive electroproduction of jets.  If the
assumption that hard PQCD dominates at each stage of the interaction
were correct, nuclear shadowing should be numerically small and
suppressed by a power of $Q^2$. The discussion in section~6 indicates
that, on the contrary, in QCD at sufficiently small $x$ and fixed
$Q^2$ nuclear shadowing is expected to be substantial and universal
for all hard processes. This conclusion is supported by current data
on nuclear shadowing in deep inelastic processes.

Dijet production has been also considered in the constituent quark
model of the proton~\cite{NZ,NZZ}. In this approach the cross section
for diffraction is expressed through a convolution of the quark
distribution in the virtual photon, the distribution of constituent
quarks in the proton and their interaction cross section. A later
generalization of this model~\cite{NZZ} includes the gluon field of
constituent quarks. In QCD though, hard processes have to be expressed
in terms of bare partons and not constituent ones. This is due to the
use of completeness of the intermediate hadronic states in hard
processes.

Equation~(\ref{eq:34d1}) implies that in this higher twist effect the
contribution of large $b$, that is of the nonperturbative QCD, is
enhanced as compared to the large $b$ contribution to the total cross section.
This result has been anticipated in the pre-QCD times~\cite{BjKogut}
and has been confirmed in QCD~\cite{FS88}. A similar conclusion has
been reached in the constituent quark model~\cite{NZZ} approach even
though it ignores the increase of parton distributions at small $b$
characteristic for QCD (see discussion in section 11).
In QCD the hard contribution is expected to become dominant only
at rather small $x$ and large $Q^2$. A similar conclusion has been
reached
for the
cross section of diffractive processes, calculated in the
approximation of the BFKL Pomeron~\cite{Bartels}, in the triple
Reggeon region when the mass of the produced hadronic system is
sufficiently large $M^2\gg Q^2$.

Note that PQCD diagrams which were found to dominate in the large mass
diffraction~\cite{Bartels} are different from those expected from the
naive application of the BFKL Pomeron~\cite{Ryskin2,NZZ} and lead to
different formulae.

To calculate this process within the more conventional leading
$\alpha_s \ln Q^2$ approximation it is necessary to realize that in
the kinematical region where $M_{q\bar q}^2\sim Q^2$ the fractions of
nucleon momentum carried by the exchanged gluons are strongly
different, $x_{\rm hard} \simeq 2x$ but $x_{\rm soft}\ll x$. This is
qualitatively different from the case of the vector meson production
considered in section 4 in which the two values of $x$ of the gluons
were comparable. This is because in the case of dijet production the
masses of the intermediate states are approximately equal to the mass
of the final state.  As a result of the asymmetry of the two $x$
values the overlap integral between the parton wave functions of the
initial and final protons cannot be expressed directly through the
gluon distribution in the target.  However at sufficiently small $x$
and large $Q^2$, when the parameter ${\alpha_s\over \pi}\ln x
\ln{Q^2\over \Lambda^2} \sim 1$, electroproduction of high $p_t$
dijets can be expressed through the gluon distribution in a target but
in a more complex way. In this particular case the factorization
theorem can be applied after the first two hard rungs attached to the
photon line, which have to be calculated exactly.  The lower part of
the diagram can be then expressed through the gluon distribution in
the target since the asymmetry between the gluons becomes unimportant
in the softer blob. The proof is the same as for the vector meson
electroproduction
\footnotemark[9]
\footnotetext[9]{ We are indebted to A.Mueller  for the discussion of
this problem.}. The cross section is proportional to
\begin{equation}
\left. {d\sigma^{\gamma^*+N \rightarrow jet_1 + jet_2 +N}\over dt}
\right| _{t=0}
\propto \left|A_{\gamma^*+gg \rightarrow jet_1+jet_2}\right|^2
\left| \tilde x  G_N(\tilde x, Q^2)\right|^2
\propto \left( {\alpha_s(Q^2)xG_N(x,Q^2) \over Q^2}\right)^2  \ ,
\label{eq:35c}
\end{equation}
where $\tilde x$ is the average $x$ of the gluons in the $\gamma^*+gg
\rightarrow jet_1+jet_2$ amplitude, $\tilde x\gg x$, and
$A_{\gamma^*+gg \rightarrow jet_1+jet_2}$ is the hard scattering
amplitude (which includes 2 hard rungs) calculated in PQCD.

 One of the nontrivial predictions of QCD is that the decomposition of
 the cross section for a longitudinally polarized photon in powers of
 $Q^2$ becomes inefficient at small $x$. This is because additional
 powers of $1/Q^2$ are compensated to a large extent by the increase
 with $Q^2$ of $\left[\alpha_s(Q^2)xG(x,Q^2)\right]^2 \sim {Q\over x}$
(see figure 4 and equations~(\ref{eq:35c}), (\ref{eq:34d1})).  Thus
the prediction of QCD is that electroproduction of hadron states with
$M_{X}^2\ll Q^2$ by longitudinally polarized photons, formally a higher
twist effect, should in practice depend on $Q^2$ rather mildly.  The
contribution of such higher twist effects to the total cross section
for diffractive processes may be considerable, as high as $ 30-40 \%
$.  One of the observed channels, the electroproduction of $\rho$
mesons, constitutes probably up to $10 \% $ of the total cross section
for diffractive processes. So far a detailed quantitative analysis of
this important issue is missing. On the experimental side, it would be
extremely important to separate the longitudinal and transverse
contributions to diffraction.
\begin{figure}
\centerline{\epsffile{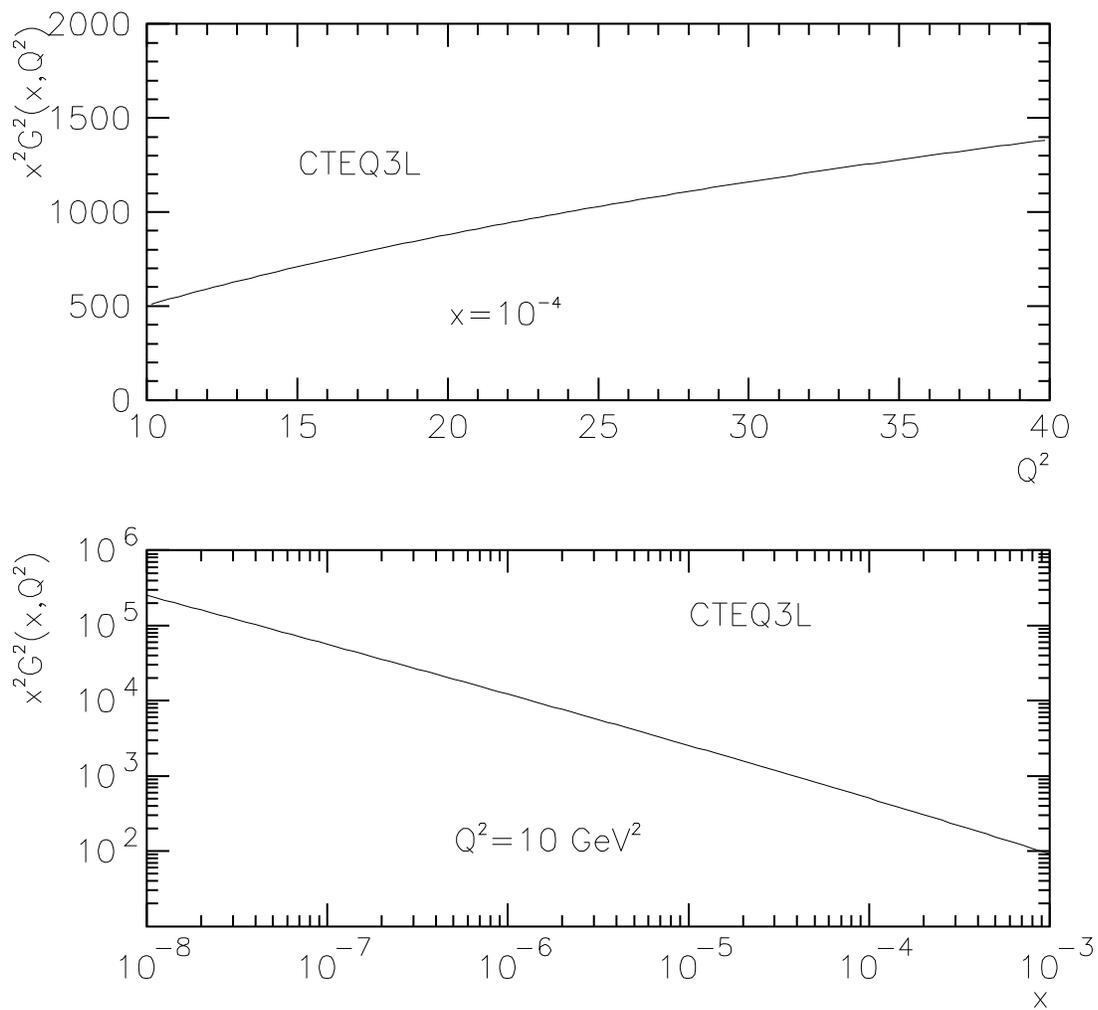}}
\vspace{-5cm}
\protect{\caption{The square of the $xG(x,Q^2)$ distribution as a
    function of $Q^2$ for a fixed $x=10^{-4}$ and as a function of $x$
    for a fixed $Q^2=10$~GeV$^2$ for the CTEQ3L parameterization.}}
\end{figure}

\section{Limiting behavior of cross sections for hard processes }

Both the GLDAP QCD evolution equation~\cite{GLDAP} and the BFKL
equation~\cite{BFKL} predict a significant increase of parton
distributions at large $Q^2$ and small $x$. This expectation is
consistent with a fast increase of parton densities with decreasing
$x$ observed at HERA. The question then arises whether the increase of
parton distributions with increasing $Q^2$ and ${1\over x}$ will stop
or will continue forever.  The evident lack of elastic unitarity
condition for the electromagnetic
amplitude precludes the use of theoretical approaches which lead to Froissart
limit for the on mass shell amplitude. (The elastic contribution to the
unitarity condition is suppressed by power of the electromagnetic
coupling constant.) At the same time it is well known that the PQCD
calculation of the scattering of two body amplitude via the sum of
leading logarithms runs, at sufficiently large energies, into conflict
with s-channel unitarity -- violation of the Froissart limit.  How to
restore unitarity still remains an open question. Thus the leading
logarithms approximation is expected to be applicable only in a
restricted kinematical region.

Our aim here is to estimate the kinematical region of applicability of
the leading logarithms approximation from the requirement of
self--consistency of QCD calculations of the hard diffractive
processes discussed in sections 4--9. We will also consider the
unitarity condition for the scattering of colorless wave packets to
deduce restrictions on the limiting behavior of cross sections for
deep inelastic processes. The idea is rather simple. In QCD, within
the leading $\alpha_s\ln Q^2$ and/or $\alpha_s \ln x$ approximation
$\sigma_L^{tot} (\gamma^* +p)$ should increase at small $x$ like
$xG_p(x,Q^2)$. At the same time the cross sections for diffractive
electroproduction of states with $M_X^2 \ll Q^2$, $\sigma_L(\gamma^*
+p\rightarrow X +p)$, is proportional to $[xG_p(x,Q^2)]^2$ as
discussed in section 4.  Thus the requirement that $\sigma_L^{tot}
(\gamma^* +p) \gg \sigma_L(\gamma^* +p\rightarrow X +p)$ will lead to
a restriction on the region of applicability of the leading logarithm
formulae. We do not have as yet measurements of the total cross
section for longitudinal virtual photons, but we can estimate in which
range of $x$ the contribution from small mass diffractive production
will be comparable in size to the contribution expected from the QCD
evolution equation. For convenience we will introduce the ratio $R_L$
defined as follows: $$R_L={\sigma_L(\gamma^* +p\rightarrow X +p)\over
  \sigma^{tot}_L (\gamma^*+p)} \ .$$ Requiring $R_L\leq 1$ and
assuming that for $Q^2=10$~GeV$^2$
 \begin{equation}
\sigma_L^{tot}(\gamma^* +p) \leq 0.5 \,
\sigma_T^{tot}(\gamma^* +p)
\end{equation}
the inconsistency of the leading $\alpha_S \ln x$ approximation and
the evolution equation should occur at $x\leq 10^{-6}$.  In practice
one should require a more severe limit on $R_L$, since the diffraction
cross section is always a small part of the total cross section
especially in the limit of black interactions. Let us take for
illustration $R_L=0.2$ as the limit. In this case the slowing down of
structure functions increase should occur for $x\sim \leq 10^{-4.3}$.

There is a certain similarity between these estimates and the
estimates exploring the increase with $\frac{1}{x}$ of shadowing
corrections to the QCD evolution equation which were calculated within
the BFKL approximation. In~\cite{LR} shadowing corrections were
estimated within the constituent quark model with the radius of the
constituent quark
equal to
$1\over 2.5$~GeV$^{-1}$. In QCD the value of shadowing
corrections calculated in~\cite{MQ} by iterating the hard amplitude
depends strongly on the fitting parameter -- the correlation radius of
gluons in a average configuration in the wave function of a hadron.
Theoretical calculations made in~\cite{MQ} found that for realistic
parameters, corresponding to a correlation radius comparable to the
radius of a hadron, the value of hard shadowing corrections to parton
distributions is negligible in the kinematical range of HERA. (For
recent numerical calculations see~\cite{BKCK,Martin}).  The major
difference between our estimate and the one implied by shadowing
corrections is that we consider cross sections of coherent diffractive
dissociation into small masses which includes the nonperturbative QCD
effects in a different and rather reliable way.  The derivation of the
region of applicability of the leading $\alpha_s\ln x$ approximation
and/or evolution equation for the cross section of longitudinally
polarized photons is based on equations which use experimentally
measurable quantities.

An estimate of where the consistency of QCD expectations fails can also be
obtained from the consideration of the ratio of cross sections
for the diffractive production of small masses and for the inclusive
diffractive production $\sigma_D(\gamma^* +p)$,
$$ R_D = \frac{\sigma_L(\gamma^* +p\rightarrow X +p)}{\sigma_D
(\gamma^* +p)} \ .$$ The ratio $R_D$ should be smaller than 1. From the
discussion presented  in this lecture we expect,
$$R_D \propto {[xG_p(x,Q^2)]^2 \over Q^4F_{2p}(x,Q^2)} \ .$$
Assuming that $R_D =~0.1$ at $Q^2 = 10$~GeV$^2$ and $x=10^{-3}$
and using the new CTEQ3L parameterization~\cite{CTEQnew}
we obtain $R_D=1$ for $x \sim 10^{-5.5}$
  (see figure 5).
\begin{figure}
\centerline{\epsffile{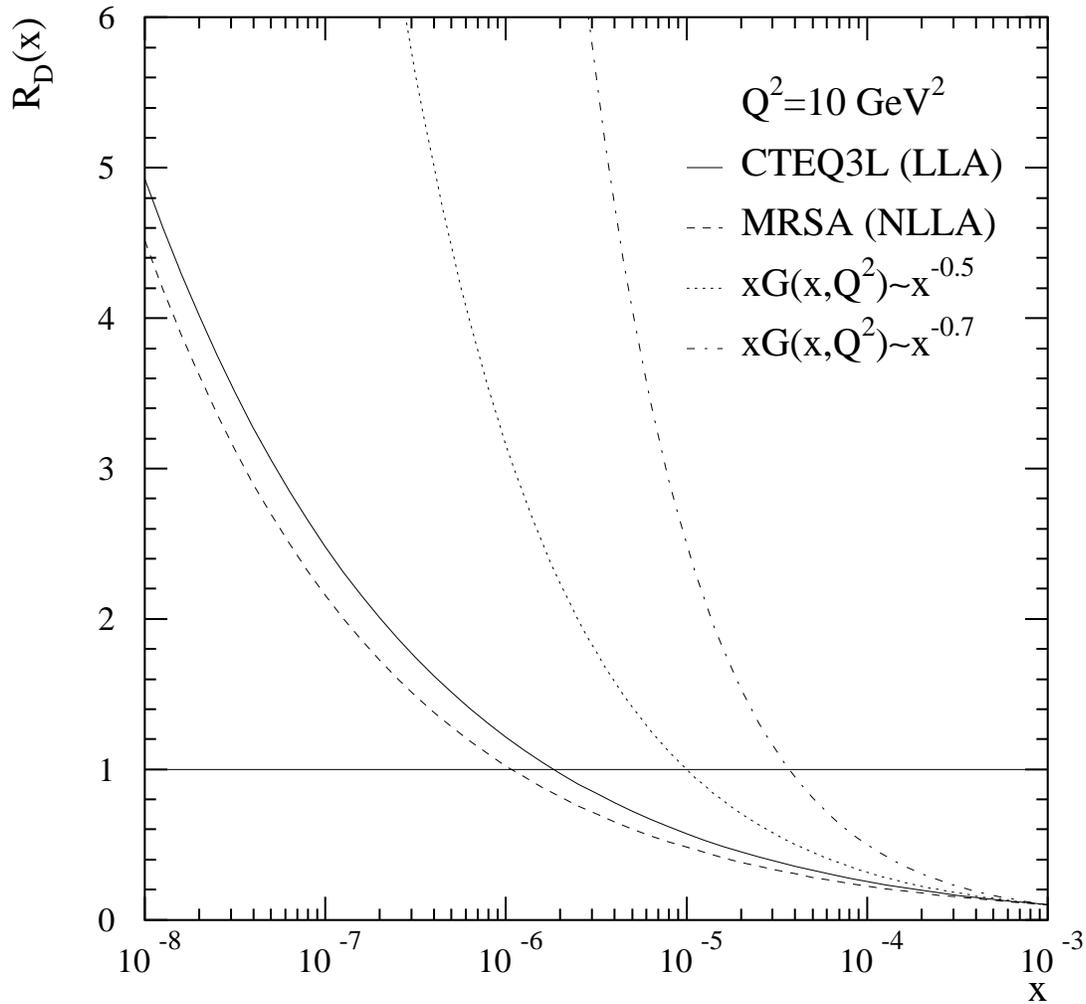}}
\vspace{-5cm}
\protect{\caption{The ratio $R_D \sim \frac{(xG(x,Q^2))^2}{Q^4F_2}$
normalized to be 0.1 at $x=10^{-3}$ at $Q^2=10$~GeV$^2$ for two
different parameterizations of parton distributions and obtained
assuming $xG(x,Q^2) \sim x^n$ as indicated in the figure.}}
\end{figure}

If one were to use the BFKL prediction for the rise of gluons (with an
intercept of the BFKL Pomeron of 1.7 at $Q^2 = 10$~GeV$^2$~\footnote{In
the expression for the intercept of the singularity corresponding to the BFKL
Pomeron, $n=1+{12\alpha_s\ln2\over \pi}$, we chose $\alpha_s(Q^2=10~{\rm
GeV}^2)=0.25$ which leads to $n(Q^2=10{\rm~GeV}^2)=1.7$.}),
which is faster than that of the CTEQ parameterization, the result
would be that $R_D=1$ at $x=10^{-4.4}$.

The value $R_D=1$ which we used to deduce this restriction is
unrealistically large since $\rho$ meson production is one of many
possible channels for the fragmentation of a $q\bar q$ pair.  We thus
expect that the fast increase of the cross section for hard
diffractive processes should stop at much larger $x$.  Assuming a
plausible value for $R_D$, $R_D=0.4$, the slow down should occur
around $x \sim 10^{-4}$ or even earlier. Note that in QCD the cross
sections for hard diffractive production of states X with $M_X^2\ll Q^2$
should have the same $x$ dependence, independent of $M_X$.

It is also interesting to note that we overestimate the value of $1/x$
since we ignore other higher twist diffractive processes which
decrease slower with $Q^2$ (cf. discussion in section 10).  The
contribution of hard diffraction dissociation into large masses $M^2$
($Q^2\leq M^2$) has been also ignored in the above analysis.  Hard
diffraction dissociation into large masses has been recently estimated
in PQCD in~\cite{Bartels} in the triple Pomeron limit. Including this
term would lead to a more stringent restriction on the region of
applicability of QCD evolution equation and/or leading $\alpha_s \ln
x$ approximation. Thus our conclusion is that violation of leading
logarithms approximation is expected in a range of $x$ which is in the
reach of accelerators such as HERA or LHC.

We shall now deduce a more stringent restriction on the increase of
parton distributions based on the interaction picture in the
laboratory frame.  We consider here the scattering of a small object,
a $q\bar q$ pair, from a large object, a nucleon. If only hard physics
was  relevant for the increase of parton distributions at small $x$,
the radius of a nucleon should not increase (small Gribov diffusion).
Under this assumption the unitarity limit corresponds to a black
nucleon.  In this case the inelastic cross section cannot exceed the
geometrical size of the nucleon
 \begin{equation}
\sigma(q\bar q N)= \frac{\pi^2}{3} b^2\alpha_s(1/b^2)
xG_N(x,b^2) <\pi r_N^2 \ .
\label{60d}
\end{equation}
To find the value of $r_N$ in eq.(\ref{60d}) we
 use the optical
theorem to calculate the elastic cross section for a $q\bar q$ pair
scattering off a nucleon,
\begin{equation}
\sigma_{el}={\sigma_{tot}^2\over 16\pi B}
\label{61d}
\end{equation}
where $B$ is the slope of the elastic amplitude (cf. discussion in
section 4). To deduce the inequality (\ref{60d}) we assume that the
interaction of a $q\bar q$ pair with a target is black and therefore
the elastic cross section is equal to the inelastic cross section
$\sigma_{el} \le \sigma_{inel}$. Based on this we find $r^2_N=4 B
\simeq 16{\rm~GeV}^{-2} \simeq (0.8 {\rm~fm})^2$ is the radius of a nucleon.
It follows from the above equations that practically the same estimate
is obtained from the assumption that ${\sigma(el)\over\sigma(tot)}\sim
(0.3-0.5)$.

According to our estimate (in sections 4 and 5) for
$Q^2=10$~GeV$^2$,~~ $b=0.3{\rm~fm}$ at least for higher twist effects.
If we then determine at which $x$ the geometrical cross section is
saturated we obtain for $b=0.3 {\rm~fm}$ and
$\alpha_s(Q^2=10{\rm~GeV^2})=0.3$, $xG_p(x,Q^2=10{\rm~GeV^2}) \le 43$.
If for illustration we use a parameterization $xG_p(x,Q^2)=4/3
x^{-0.36}$ which
is  consistent with
the current data we find that $x \ge 6.0~ 10^{-5}$. Similar consideration
for  the case of scattering off nuclei leads to
\begin{equation}
{1 \over A} xG_A(x,Q^2=10{\rm~GeV^2}) \le {43 \over A^{1/3}}
\label{AA}
\end{equation}
Restriction (\ref{AA}) has obvious practical implications for nucleus-nucleus
collisions at LHC energies. In particular, it strongly affects theoretical
predictions of the
cross section of the minijet, and heavy quark production say for
 $Q^2=10 GeV^2$ and $x \le 10^{-3}$.

The use of the amplitude for $q\bar q$ pair scattering off a nucleon
to deduce the limit allows to account accurately for
nonperturbative QCD effects through the unitarity condition for such
an amplitude.  On the other hand if the increase of parton
distributions is related to soft physics as well then the cross
section may be allowed to increase up to smaller $x$ values.

The black disc limit for $\sigma_{\gamma^*N}$ has been discussed
earlier (for a review and references see~\cite{BKCK,Levin}). The
difference compared to previous attempts is that we deduce the QCD
formulae for the cross section of a $q\bar q$ pair scattering off a
hadron target.  For this cross section the geometrical limit including
numerical coefficients unambiguously follows from unitarity
of the $S$-matrix,
that is
the geometry of the collision. As a result we obtain an inequality
which contains no free parameters.  Recently a quantitative estimate
of the saturation limit was obtained~\cite{Collins} by considering the
GLR model~\cite{LR,MQ} of nonlinear effects in the parton evolution
and requiring that the nonlinear term should be smaller than the
linear term.  The constraint obtained for $xG_p(x,Q^2)$ is numerically
much less restrictive compared  to our result.
Even a more stringent restriction follows
for the interaction of a colorless gluon pair off a nucleon from the
requirement that the inelastic cross section for the scattering of a
small size gluon pair should not exceed the elastic one
\begin{equation}
\sigma(gg N)= \frac{3\pi^2}{4} b^2\alpha_s(1/b^2)
xG_N(x,b^2) <\pi r_N^2 \ .
\label{60f}
\end{equation}
For $b=0.25{\rm~fm}$ the geometrical limit is achieved for $x\sim
10^{-3}$.

We want to point out that the black disc limit implies a restriction
on the limiting behavior of the cross sections for hard processes but
does not allow to calculate it. The dynamical mechanism responsible
for slowing down of the increase of parton distributions so that they
satisfy equations~(\ref{60d},~\ref{60f}) is not clear.  In particular
the triple Pomeron mechanism for shadowing suggested in~\cite{LR} does
not lead to large effects at HERA energies especially if one assumes a
homogeneous transverse density of gluons~\cite{Levin,Martin}.

The theoretical analysis performed in this section does not allow to
deduce restrictions on the limiting behavior of parton distributions
in a hadron. Beyond the evolution equation approximation and/or
leading $\alpha_s \ln x$ approximation the restriction on the cross
sections of deep inelastic processes cannot be simply expressed in
terms of parton distributions in a hadron target.

We want to draw attention to the fact that nonperturbative QCD effects
play an important role in the contribution of higher twist effects to
$\sigma_L(\gamma^* p)$.  This is evident from the impact parameter
representation of the contribution to $\sigma_L(\gamma^* +p)$ of $n$
consecutive rescatterings of small transverse size $q\bar q$
pairs.
 This contribution is proportional to
$$Q^2\int |\psi_{\gamma_L}^*(z,b^2)|^2 dz  d^2b
\left[\alpha_s(1/b^2)b^2xG(x,b)\right]^{n} \ .$$ The
  inspection of this integral shows that for large $n\geq 3$, $b$ which
  dominates under the integral does not decrease with increasing $Q^2$
for $x \sim 10^{-3} \div 10^{-4}$.
We use as estimate $xG_N(x,Q^2) \propto \sqrt{Q} $ which follows
from the evolution equation for small $x$.
(This QCD effect is absent in the applications~\cite{NZZ} of the
constituent quark model).
Thus if higher twist effects were really important in small $x$
physics, it would imply that the small $x$ physics is the
outcome of an interplay of hard (small $b$) and soft (large $b$) QCD.
To illustrate this point let us consider the cross section of
diffractive electroproduction of hadrons with masses $M^2\sim Q^2$ by
transversely polarized photons. Applying the same ideas as in the case
of longitudinally polarized photons we would obtain a similar
expression as given by equation~\ref{eq:34d1}. The important
difference is that the wave function of a transversely polarized
photon is singular for $z\rightarrow 0$ or $1$. As a result the
contribution of large impact parameters $b$ in the wave function of
the photon should give the dominant contribution to the integral in a
wide kinematical range of $x$ and $Q^2$. This has been understood long
ago -- see discussion in sections 13-14.  A similar conclusion has
been achieved recently~\cite{NZZ} within the constituent quark model.
(Note however that this model ignores the increase of gluon
distribution with Q typical for QCD and therefore overestimates the
nonperturbative QCD contribution). Thus such type of diffractive
processes should depend on energy in a way similar to the usual soft
hadron processes.

A good example of the consequences of the interplay of small $b$ and
large $b$ physics is that in electroproduction of small mass states
the unitarity limit may become apparent at larger $x$
than in the case of the total cross section of deep inelastic processes.

\section{Geometrical limit.}

The important role of nonperturbative QCD effects clearly shows that
new ideas beyond PQCD approach are required.

The starting point of our discussion is the observation that when
$l_c$ considerably exceeds the diameter of a target which is at rest,
the virtual photon transforms into hadron components well before the
target. Thus small $x$ physics probes the interaction of various
hadron wave packets with a target (for the sake of the argument we
will consider the interaction with a proton target). The geometrical
limit for the cross section of virtual photon scattering off a nucleon
target will follow from the assumption that a target is black for the
dominant hadron components in the wave function of the virtual photon.
In this approximation, the structure functions of a target can be
unambiguously calculated.  A comparison with the formula deduced from
the geometrical limit will imply certain restrictions on the increase
of structure functions~\cite{Bj94}.

Blackness of hadron interactions at high energies has been predicted
in the elastic eikonal models with input cross sections increasing
with energy (see~\cite{CW} and references therein). The assumption on
the blackness of hadron-hadron interactions at high energies is often
used to deduce the Froissart limit.

Under the assumption that the interaction is black,
the cross section of non-diagonal diffractive processes should be small
and decrease with increase of energy.  This has been understood long
ago by considering the example of an energetic particle scattering off
a black potential well (see e.g.~\cite{Landau}).

Within this picture, in the limit $x\rightarrow 0$, the deep inelastic
lepton scattering off a nucleon can be unambiguously calculated
through the cross section of the reaction $e^+e^-\rightarrow
$~hadrons~\cite{Gribov69},
\begin{equation}
\sigma_{inel} (\gamma^* +p)=
{\alpha\over 3 \pi}\left<\sigma\right>
\int\limits^{M^2_{max}}
_{m_0^2} R(m^2)
{m^2 \over (m^2 +Q^2)^2} dm^2 \ ,
\label{eq:38c}
\end{equation}
where $\left<\sigma\right>$ is the average interaction cross section,
$m$ denotes the mass of the state contributing to the deep
inelastic cross section and at the same time the center
of mass energy of the $e^+e^-$ system,
$m_0^2 $ is the hadronic scale $\sim 1 $~GeV$^2$ and $R$ is defined by
\begin{equation}
R(m^2) ={{\sigma (e^+e^-  \rightarrow {\rm hadrons} ) }\over
\sigma (e^+e^-  \rightarrow
\mu^+ \mu^- )} \ .
\label{eq:39c}
\end{equation}
The upper limit in
formula~(\ref{eq:38c}), $M^2_{max} \sim Q^2
({1\over mr_N} -1)$ with $r_N$ the radius of a nucleon,
follows from the condition that the
diffractive production of a state with mass $m$ should not be suppressed
by the nucleon form factor. Thus
\begin{equation}
\sigma (\gamma^* +p)= P_{had}\left<\sigma\right> \ ,
\label{eq:40c}
\end{equation}
where the probability of a given configuration $P_{had}$ is expressed by
\begin{equation}
P_{had}={\alpha\over 3\pi}\int R(m^2){m^2 \over (m^2 +Q^2)^2} dm^2 \ .
\label{eq:41c}
\end{equation}
Since large masses dominate in the integral it is legitimate
to substitute $R(m^2)$ by its asymptotic value.
For numerical estimates we will use $R$ for the case of 5 flavors,
$R={11\over 3} $. Thus  in the unitarity limit
\begin{equation}
P_{had}={\alpha \over 3\pi}R \left[\ln({1\over 10x }) -1\right] \ .
\label{eq:42c}
\end{equation}
In the integral over $m^2$ we use the conventional definition
of diffraction as leading to hadronic final states with
${M_{X}^2\over s}\leq 0.1$ where diffractive production is
not suppressed by the  nucleon form factor.

The expression for $F_{2p}$ follows from equations~(\ref{eq:40c})
and~(\ref{eq:41c}) and the relation between the total $\gamma^* p$
cross section and the structure function $F_{2}$
\begin{equation}
F_{2p}(x,Q^2)={Q^2 \left<\sigma\right>
 \over 12\pi^3}R \left[\ln {1\over 10x } -1\right] \ .
 \label{eq:43c}
\end{equation}
For the proton target following~\cite{Bj94}
we estimate
\begin{equation}
\left<\sigma\right> = \sigma_{pp}/2 \ .
\label{eq:44c1}
\end{equation}
For further estimates we  shall use a parameterization of $F_{2p}(x,Q^2)$
given by expression~(\ref{eq:lc}).

Assuming that $F_{2p}$ cannot exceed the unitarity limit given
by~(\ref{eq:43c}) and taking $\left<\sigma\right>=40$~mb allows to
obtain some constraints on the region where parton distributions may
increase with Q and ${1\over x}$~\cite{Bj94}.  Assuming that the
parameterization given by~(\ref{eq:lc}) holds down to extremely low
$x$, the geometrical limit would be exceeded at $x\sim
10^{-7},~10^{-8}$ for $Q^2 = 10$~GeV$^2$. This is three orders of
magnitude below the present kinematical limit at HERA.  The
restrictions obtained from the geometrical limit
for the cross section averaged over impact parameters
are thus interesting
from a purely theoretical point of view but not for any practical
purposes.  Somewhat more useful restrictions on the small x behavior
follow from the mismatch of QCD predictions for inclusive and
exclusive processes and from $S$-matrix unitarity constraint.

\section{Diffraction in DIS at intermediate $Q^2$}

It has been understood long ago that the production of almost on mass
shell quarks by virtual photons should give a significant contribution
to the total cross section for deep inelastic scattering at small
$x$~\cite{BJ71}.  One of the predictions of this approach (which is
essentially the parton model approximation) is a large cross section
for diffractive processes. The  QCD $Q^2$ evolution does not change this
physical picture radically. The only  expected modification of the picture is
the appearance   of a number of hard jets in
the current fragmentation region~\cite{FS88} typical for
including  $\alpha_S \ln Q^2$ terms.  It is often stated   that the
dominance of the BFKL Pomeron in diffractive processes
predicts the dominance of final states consisting
of hard jets ~\cite{Ryskin2,LW}.  However this prediction is not robust since
 the analysis of Feynman diagrams for hard processes
in QCD finds strong diffusion effects into the region of small
transverse momenta of partons (see~\cite{Bartels} and references
therein).  Recent HERA data~\cite{MD} seem to support the picture with
a dominance of events with small $k_t$. Thus it seems worthwhile to
investigate the role of nonperturbative QCD physics in diffractive
processes.

The interaction of a virtual photon with a target at intermediate
$Q^2$ and small $x$, when gluon radiation is negligible, can be
considered as a transformation of $\gamma^*$ into a $q \bar q $ pair
which subsequently interacts with the target.  In this case an
important role is played by the quark configurations in which the
virtuality of the quark interacting with the target is small,
\begin{equation}
k_{qt}\sim  k_{t0} \ ,\ \ \ \alpha_q = {(m_q^2 +k_{qt}^2)\over Q^2} \ .
\label{soft}
\end{equation}
Here $\alpha_q$ denotes the light-cone fraction of the photon momentum
carried by the slower quark and $k_{t0}$ is an average transverse
momentum of partons in the hadron wave function, typically $k_{t0}
\sim 0.3-0.4 $~GeV.

In the language of non-covariant perturbation theory the $q\bar q$
configurations described by~(\ref{soft}) correspond to an intermediate
state of mass $m^2 \sim Q^2$ and of transverse size $\sim {1 \over
  k_{t0} } \geq 0.5$ fm.  These configurations constitute a tiny
fraction $\sim {k_{qt}^2\over Q^2}$ of the phase volume kinematically
allowed for the $q \bar q$ pair.  However the interaction in this case
is strong -- similar to the interaction of ordinary hadrons, since the
virtuality of the slower quark is small and the transverse area
occupied by color is large.  The contribution of these configurations
leads to Bjorken scaling since the total cross section is proportional
to ${1\over Q^2}$ and in the parton model only these configurations
may contribute to the cross section. Hence Bjorken has
assumed~\cite{BJ71} that all other configurations are not important in
the interaction though the underlying dynamics of such a suppression
was not clear at that time~\cite{BjKogut}.  Accounting for the $
({k_{qt}^2\over Q^2})$ factor in equation~(\ref{eq:38c}) allowed him
to reconcile the Gribov dispersion representation with scaling.  He
suggested to refer to these configurations as aligned jets since both
quarks have small transverse momenta relative to the photon momentum
direction. In further discussions we will refer to this approach as
that of the Aligned Jet Model (AJM).  Note that in terms of the
Feynman fusion diagram the aligned jet contribution arises only for
transversely polarized virtual photons. This is because the vertex for
the transition $\gamma^*_T \rightarrow q \bar q$ is singular $ \sim
{1\over z}$ when the fraction of the photon momentum $z$ carried by
the slowest quark (antiquark) tends to 0. For the case of a
longitudinally polarized photon the aligned jet approximation produces
results qualitatively different from expectations in QCD where the
contribution of symmetric jets dominates. This is because in QCD the
dominant contribution to the $\gamma^*_L$--nucleon cross section
arises from the region of large $k_{qt} \sim {Q\over 2}$.

In QCD the interaction of quarks with large relative transverse
momenta with a target is suppressed but not negligible. The
suppression mechanism is due to color screening since $q \bar q$
configurations with large $k_t$ correspond, in the coordinate space,
to configurations of small transverse size, $b \sim {1\over k_t}$, for
which equation~(\ref{eq:9c}) is applicable. It is easy to check that
the contribution of large $k_t$ also gives a scaling contribution to
the cross section. The practical question then is which of the two
contributions dominates at intermediate $Q^2 = Q_0^2 \approx 4
$~GeV$^2$, above which one can use the QCD evolution equations. To
make a numerical estimate we assume that the $q \bar q$ configurations
with $k_{qt} \leq k_{t0}$, in which color is distributed over a
transverse area similar to the one occupied by color in mesons,
interact with a cross section similar to that of the pion. A
comparison with experimental data for $F_{2p}(x \sim 0.01, Q^2_0) $
indicates that at least half of the cross section is due to soft, low
$k_t$ interactions~\cite{FS88,FS89}.  A crucial check is provided by
applying the same reasoning to scattering off nuclei in which the
interaction of the soft component should be shadowed with an intensity
comparable to that of pion-nucleus interaction.  Indeed the current
deep inelastic data on shadowing for $F_{2A}(x,Q^2)$ are in
reasonable agreement with calculations based on the soft mechanism of
nuclear shadowing~\cite{FS88,FS91}.

Similarly to the case of hadron-nucleon and hadron-nucleus
interactions, the interaction of $\gamma^*$ in a soft hadron component
naturally leads to diffractive phenomena. Application of the Gribov
representation with a cutoff on the $k_t$ of the aligned jets in the
integral leads to a diffractive mass spectrum for the transversely
polarized virtual photon~\cite{BjKogut}
\begin{equation}
{d \sigma \over d M^2} \propto {1\over (M^2+Q^2)^2} \ .
\label{dajm}
\end{equation}
The two major differences compared to the hadronic case are that
elastic scattering is substituted by production of states with $\left<
M^2\right> \approx Q^2$
and that the contribution of configurations of small spatial size
is larger for $\gamma^*_L$.

If the aligned jet configurations were dominant, the fraction of cross
section of deep inelastic $\gamma^*N$ scattering due to single
diffractive processes would be
\begin{equation}
R_{\rm single~dif}^{\rm AJM}= \sigma_{\rm dif}/\sigma_{\rm tot}=
  {\sigma_{\pi N}({\rm el})+\sigma_{\pi N}({\rm dif})\over
\sigma_{\pi N}({\rm tot})} \sim 0.25 \ \ .
\label{eq:43c1}
\end{equation}
Our numerical estimates indicate that for $Q^2 \sim Q^2_0$ and $ x
\sim 10^{-2}$ the AJM contributes about $\eta \sim 60-70 \%$ of the
total cross section. So we expect that in this $Q^2$ range the
probability for diffraction is
\begin{equation}
R_{\rm single~dif}= \eta R_{\rm single~dif}^{\rm AJM} \sim 15\%.
\end{equation}
This probability is actually related in a rather direct way to the
amount of shadowing in interactions with nuclei in the same kinematic
regime, so it is quite well determined by the nuclear shadowing data.

To estimate the probability of events with large rapidity gaps one has
to add the processes of diffractive dissociation of the nucleon and
double diffraction dissociation, leading to an estimate
\begin{equation}
P_{\rm gap} =(1.3-1.5) R_{\rm single~diff} \sim 0.2 \ .
\label{44}
\end{equation}
This is rather close to the observed gap survival probability for
photoproduction processes~\cite{photogap}.

The characteristic features of the AJM contribution which can be
checked experimentally are the charge and flavor correlations between
the fastest and the slowest diffractively produced hadrons which
should be similar to those in $e^+e^-\rightarrow$~hadrons at $M^2\sim
Q^2$.

Another important feature of the soft contribution which distinguishes
it from the contribution of hard processes is the $t$ dependence of
the cross section for $M^2 \le Q^2$. Since the size of the
configurations is comparable to that of the pion one may expect that
the $t$ slope of the cross section, $B$, should be similar to that of
the pion-nucleon interaction, i.e. $B \ge 10$~GeV$^{-2}$ which is
much softer than for hard processes where we expect $B \approx
4$~GeV$^{-2}$ (see discussion in section 4).  The large value of
the slope for the soft component is also natural in the parton type
logic where only slow partons interact. It is easy to check that for
$-t \gg k_{t0}^2 \sim 0.1$~GeV$^2$ the mass of the produced hadron
system is larger than the mass of the intermediate state by factor
${\sqrt{-t} \over k_{t0}}$. Thus for large $t$ the production of
masses $M \le Q$ is suppressed. Therefore the study of the
$t$-dependence of diffraction can be used to disentangle the
contribution of soft and hard mechanisms.

This discussion indicates also that the contribution of non-diagonal
transitions $"M^2" \rightarrow "M^{\prime~2}$ leads to a weaker decrease
of the differential cross section with $M^2$ than given by
equation~(\ref{dajm}). Besides at large $M^2 \sim ~few ~Q^2$ one expects
an onset of the dominance of the triple Pomeron mechanism which corresponds
to
\begin{equation}
{d \sigma \over d M^2} \propto {1 \over Q^2 M^2}.
\label{PPP}
\end{equation}
\section{$Q^2$ evolution of the soft contribution
in diffraction.}

The major difference between the parton model and QCD is the existence
in QCD of a significant high $p_t$ tail in the parton wave functions
of the virtual photon and the proton. This is the source of the
violation of Bjorken scaling observed at small $x$.  It is thus
necessary to modify the aligned jet model to account for the hard QCD
physics.

It is in general difficult to obtain with significant probability a
rapidity gap in hard processes in perturbative physics.  Confinement
of quarks and gluons means that a gap in rapidity is filled by gluon
radiation in PQCD and subsequently by hadrons~\cite{Feynman71}.  It is
possible to produce diffraction in perturbative QCD but the price is a
suppression by powers of the coupling constant $\alpha_s$ and/or
powers of $Q^2$.  In first approximation in calculating diffraction in
deep inelastic processes at small $x$ we will thus neglect diffraction
in PQCD. In the following analysis, for the description of large
rapidity gap events, we shall use the QCD modification of the AJM
model suggested in~\cite{FS88} as well as the suggestion of Yu.
Dokshitzer~\cite{Doc94} to add to the conventional evolution equation
the assumption of local duality in rapidity space between quark-gluon
and hadron degrees of freedom.

In the course of the following considerations it will be convenient to
switch to the Breit frame.  In this frame the photon has momentum
$(0,-2xP)$ and the initial proton has momentum ($P$,$P$).
Correspondingly $Q^2=4x^2P^2$. The process of diffraction can be
viewed as the virtual photon scattering off a color singlet $q \bar q$
pair with the interacting parton carrying a light-cone fraction
$\alpha$ and the spectator parton carrying a light-cone fraction
$x_1$. We assume here the local correspondence in rapidity space
between partons and hadrons. The mass of the produced system, $M$ is
given by
\begin{eqnarray}
M^2=(p_{\gamma^*}+p_{x_1}+p_{\alpha})^2=
P^2((\alpha+{x_1} )^2-(\alpha+{x_1}-2x )^2)= \nonumber \\
Q^2+4P^2 (\alpha + x_1)x= Q^2 {\alpha + x_1 -x \over x} \ .
\label{B1}
\end{eqnarray}
In the approximation that gluon radiation is neglected (parton model)
$\alpha=x$ and the mass of the diffractively produced system $M$ is
\begin{equation}
M^2=Q^2 x_1/x \ .
 \label{B2}
\end{equation}
The differential cross section for production of mass $M$ follows from
equation~(\ref{dajm}),
\begin{eqnarray}
{d \sigma^{\rm AJM}\over d  M^2} =
\Gamma \int d x_1 \delta (x_1 -{x M^2\over Q^2}){1 \over  (Q^2
+M^2)^2}= \nonumber \\
{\Gamma
\over Q^4} \int d x_1 \delta(x - \alpha)
\delta
 (x_1 - {x M^2 \over Q^2})
{1 \over (1 + x_1/x)^2} \ .
\label{B3}
\end{eqnarray}
Here $\Gamma$ is the factor which includes the density of correlated
color singlet pairs and the cross section for interaction of the
photon with the parton.  The total cross section for diffractive
dissociation comes out to be proportional to ${1 \over Q^2}$,
\begin{equation}
\int {d\sigma^{AJM} \over dM^2}
 d M^2={\Gamma \over Q^2}\int
 {d x_1 \over x}
{1 \over (1 + x_1/x)^2} = {\Gamma
\over
 Q^2} \ .
\label{B4}
\end{equation}
We do not restrict the integration over $x_1$ in equation~(\ref{B4})
since the major contribution comes from the region of $x_1\sim x$.
Thus we can formulate diffraction in the infinite momentum frame as a
manifestation of short rapidity range color correlation between
partons in the nonperturbative parton wave function of the nucleon. To
calculate the $Q^2$ evolution in QCD we have to take into account that
the parton with momentum fraction $\alpha$ has its own structure at
higher $Q^2$ resolution and that the $\gamma^*$ scatters off
constituents of the "parent" parton. This is the usual evolution with
$Q^2$ which can be accounted for in the same way as in the QCD
evolution equations by the substitution
$$\Gamma \delta (x-\alpha) \rightarrow P\sum\limits_{j} e_j^2 d_j^{\rm
  pert}({x \over \alpha},Q^2,Q_0^2)$$ where $d_j^{\rm
  pert}(x,Q^2,Q_o^2)$ are the structure functions of the parent
parton.  This effect leads to the change of the relationship between
$x_1$ and $x$ resulting from parton bremsstrahlung,
\begin{eqnarray}
  {d \sigma^{\rm soft+QCD} \over dM^2}= {P\over Q^4}\int d x_1 \int
  {d\alpha \over\alpha} \sum\limits_{j} e^2_j d_j^{\rm pert}({x \over
    \alpha},Q^2,Q_0^2) d_j^{\rm nonpert}(\alpha, Q_0^2) \nonumber \\
  \delta (x_1 + \alpha -x -{x M^2 \over Q^2}){1\over x}{1 \over
    (1+x_1/x)^2} \theta (\lambda - {\alpha + x_1-x\over 1-x})\ .
\label{B5}
\end{eqnarray}
$P$ denotes the probability of diffractive scattering in a soft
interaction and $d_j^{\rm nonpert}(\alpha, Q_0^2)$ is the parton
distribution in the soft component producing diffraction (compare
discussion in the previous section).  The $\theta$ function term
reflects the condition that diffraction in the nonperturbative domain
is possible only for
\begin{equation}
0 \leq {M^2 \over s}= {Q^2 (\alpha +x_1-x) \over  x (\nu -Q^2)}
={\alpha +x_1-x \over (1-x)}\equiv   \lambda
\sim 0.05-0.1 \ \ .
\label{B6}
\end{equation}
After performing the integral over $x_1$ we can rewrite
equation~(\ref{B5}) in the form
\begin{eqnarray}
{d \sigma^{\rm soft+QCD}
\over d M^2} =
{P \over Q^4}
\int^1_x
{d\alpha \over \alpha}  \sum\limits_{j}  e^2_j
d_j^{\rm pert}({x \over \alpha},Q^2,Q^2_0)
d_j^{\rm nonpert}(\alpha, Q_0^2)\nonumber  \\
{1 \over ( 2-\alpha/ x+M^2/Q^2 )^2}
 \theta \left( \lambda (1-x) -{xM^2 \over Q^2} \right) \theta
\left( 1-\frac{\alpha}{x} +\frac{M^2}{Q^2}\right) \ .
\label{B7}
\end{eqnarray}
After integrating equation~(\ref{B7}) over the mass of the produced
system, we obtain for the total diffractive cross section
\begin{eqnarray}
  \sigma^{\rm soft+QCD}   = {P\over Q^2}
\int^1_x {d\alpha \over \alpha}
\sum\limits_{j} e_j^2 d_j^{\rm pert}({x \over \alpha},Q^2,Q_o^2) d_j^{\rm
nonpert}(\alpha, Q_0^2)\nonumber  \\
 \nonumber  \\
{x-\alpha + \lambda(1-x)\over 2x-\alpha +\lambda(1-x)}
\theta (\lambda (1-x) -(\alpha-x)) \nonumber  \\.
\label{B71}
\end{eqnarray}

\subsection{Evolution equation for diffraction.}

Let us rewrite equation~(\ref{B5}) in a form more convenient for the
application of the evolution equations.  To this end let us consider
the ratio

$$ R \equiv Q^2 {d \sigma^{\rm soft+QCD} \over d {M^2\over Q^2}} $$
and analyze the $Q^2$ evolution of the diffraction cross section at
{\bf fixed} $\kappa =M^2/Q^2$.  Neglecting the valence quark
contribution and calculating leading $\alpha_s \ln{Q^2\over Q_o^2}$
corrections to equation~(\ref{B71}) we find,
\begin{eqnarray}
{\partial\over \partial \ln Q^2} R(x,Q^2)= P
\int^1_x
{d\alpha \over \alpha}\left<e_j^2\right>\nonumber  \\
\left\{P_{q
 q}({x \over \alpha},Q^2) 2S(\alpha,Q^2)
+N_f P_{q g}({x \over \alpha},Q^2) G(\alpha,Q^2) \right \}\nonumber  \\
\theta \left(\lambda (1-x) -xk\right) \nonumber  \\
\theta \left(1 -{\alpha\over x}+k\right) \nonumber  \\
f(\kappa,\alpha,x)
 \label{B8} \ ,
\end{eqnarray}
where $\left<e_j^2\right>$ is the average quadratic electric charge of
partons and
\begin{eqnarray*}
 f(\kappa,\alpha,x)= (\kappa +2 -\alpha /x)^{-2}
\end{eqnarray*}
for $\kappa \leq 2$.  For larger $\kappa $ where the triple Pomeron
contribution is important $f \sim \kappa^{-1}$.

\subsection{Qualitative pattern of $x$ and  $Q^2$ dependence of diffraction.}

It is easy to see that equations~(\ref{B7},\ref{B71},\ref{B8}) lead to
the leading twist diffraction.  To see the pattern of the $x,Q^2$
dependence we can assume that $d_j^{\rm pert}(x,Q^2)={d\over x^n}$ and
$d_j^{\rm nonpert} (x,Q_0^2)={d_0\over x^{n_0}}$.  It follows from
equation~(\ref{B7}) that for $x\ll \lambda $ the ratio ${\sigma_{\rm
    diff} \over \sigma_{\rm tot}}$ does not depend on $x$. One can
also see that the characteristic gap interval is
\begin{eqnarray}
\Delta  y= \ln\frac{s}{Mm_p}= \ln\frac{s}{Q^2}+\ln{Q^2\over M~m_p}
=   \ln {1\over x} +
\ln({Q^2\over M~m_p}) \ .
\label{B9}
\end{eqnarray}
The second term $\ln\frac{Q^2}{M~m_p}$ increases with $Q^2$ in the
parton model, while the scaling violation tends to reduce this
increase since the mean value of $M^2/Q^2$ at fixed $x$ increases with
$Q^2$.

There are several qualitative differences between the QCD improved
soft diffraction and the parton model (AJM).

\noindent
(i) Due to QCD evolution the number of diffractively produced hard
jets and the average transverse momentum of diffractively produced
hadrons should increase with $Q^2$.

\noindent
(ii) The distribution of ${M^2\over Q^2}$ becomes broader in QCD with
increasing $Q^2$.

\noindent
(iii) While in the parton model the cross section for the interaction
of the longitudinally polarized virtual photon is a higher twist
effect, in QCD diffraction is a leading twist for any polarization of
the virtual photon. The final state in the case of longitudinally
polarized photons should contain at least 3 jets, two of them should
have large transverse momenta comparable with $Q$.

\subsection{Connection with the Ingelman-Schlein Model}
Ingelman and Schlein have suggested to treat hard diffractive
processes using the concept of parton distribution in the Pomeron
\cite{IS}.  In this approach one calculates the light-cone fraction of
the target carried by the Pomeron, $x_P$, and light-cone fractions of
the Pomeron momentum carried by quarks and gluons, $\beta$. It is
assumed that parton distributions in the Pomeron, $\beta q_P(\beta,
Q^2), \beta g_P(\beta, Q^2)$ are independent of $x_P$ and the transverse
momentum of the recoil nucleon. For the process of inclusive deep
inelastic diffraction $\beta $ is simply related to the observables,
\begin{equation}
\beta={Q^2 \over Q^2 + M_X^2}
\label{beta}
\end{equation}

The $Q^2$ evolution of the total cross section of diffraction as
considered in the previous subsections is consistent with the
expectation of the Ingelman-Schlein model (though the final states are
not necessarily the same).  The aligned jet model in this case serves
as a boundary condition defining parton distributions in the Pomeron
at intermediate $Q_0^2$ above which QCD evolution takes place.  The
aligned jet model corresponds to the quark distribution in the Pomeron
\begin{equation}
 \beta q_P(\beta, Q_0^2) \propto \beta .
\label{qbeta}
\end{equation}

It follows from the discussion in the end of section 13 that taking
into account the non-diagonal transitions in the aligned jet model and
the triple Pomeron contribution would make the distribution flatter.
A similar, rather flat, distribution is expected for gluons for these
$Q^2$.  This expectation of the aligned jet model is different from
the counting rule anzatz of \cite{IS}:
$ \beta q_P(\beta, Q_0^2) \propto (1-\beta)$.

\section{Non-universality of the  pomeron in QCD.}

Theoretical considerations of soft diffractive processes have
demonstrated that ordinary hadrons contain components of very
different interaction strength~\cite{BBFHS93,BBFS93}.  This includes
configurations which interact with cross sections much larger than the
average one and configurations which interact with very small cross
sections, described by equation~(\ref{eq:9c}) for a meson projectile.
The probability distribution to find a pion and nucleon in
configurations with a given interaction cross section $\sigma$ --
$P(\sigma)$ is presented in figure~6~\cite{BBFS93}, which also
includes the estimate of the probability of small cross sections in
the pion which is close in spirit to the analysis of the diffractive
$\rho$ meson production described in this report.
\begin{figure}
\centerline{\epsffile{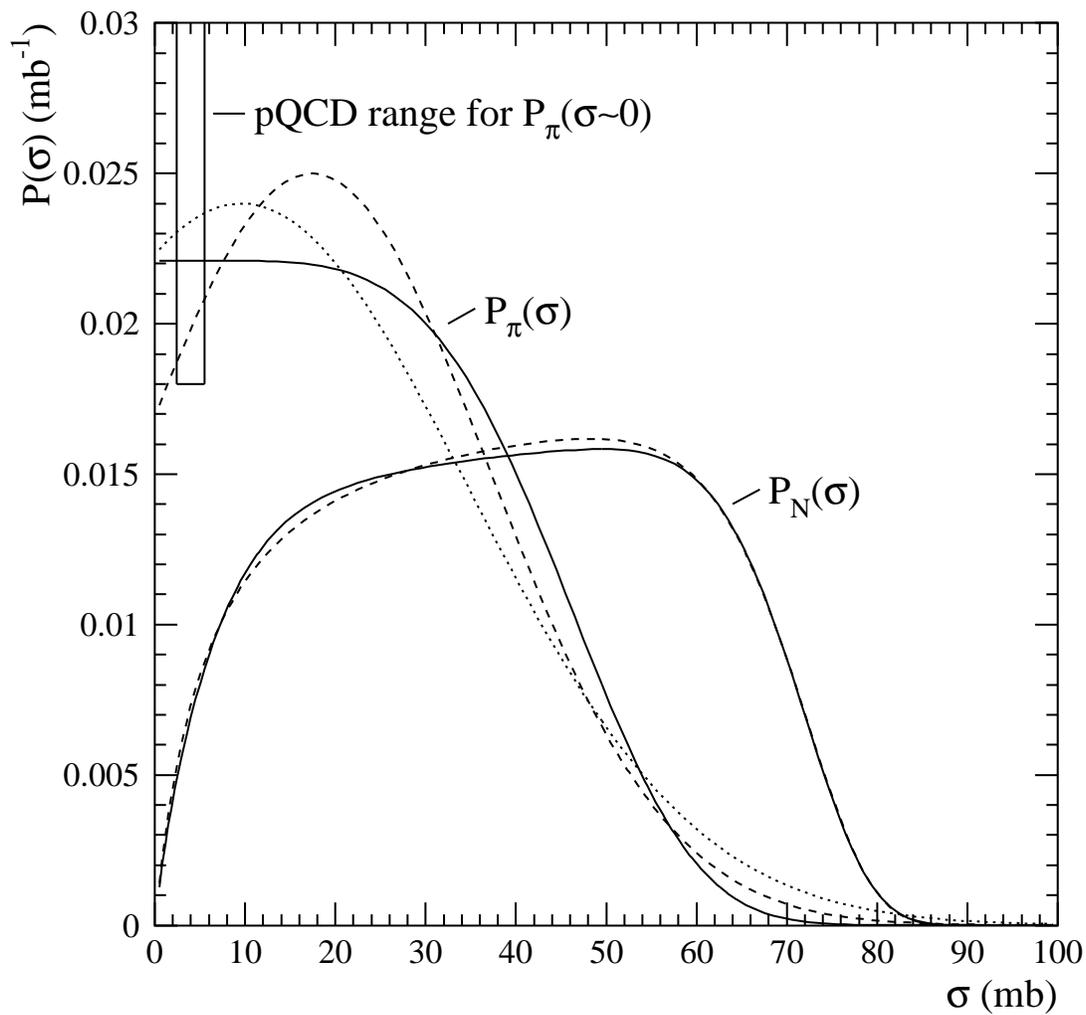}}
\vspace{-5cm}
\protect{\caption{Cross-section probability for pions
    $P_{\pi}(\sigma)$ and nucleons $P_{N}(\sigma)$ as extracted from
    experimental data.  $P_{\pi}(\sigma \sim 0)$ is compared with the
    perturbative QCD prediction.}}
\end{figure}

The presence in hadrons of various configurations of partons having
different interaction cross sections with a target is in evident
contradiction with the idea of a universal vacuum pole where universal
factorization is expected. At the same time it is well known that the
Pomeron pole approximation is not self-consistent. The vacuum pole
should be accompanied by a set of Pomeron cuts~\cite{Gribov4}.  For
the sum of the Pomeron pole and the Pomeron cuts no factorization is
expected.  Thus the S matrix description and the QCD description are
not in variance.  We shall enumerate now where and how to search for
the non universality of the effective Pomeron understood as the sum of
the pomeron pole and the Pomeron cuts.

It is natural to distinguish two basic manifestations of the
non universality of the effective Pomeron trajectory, $\alpha_{\cal
  P}(t) \approx \alpha_0 + \alpha't$, a different energy dependence of
the interaction cross section, which is characterized by a different
value of $\alpha_0$, and a different rate of the Gribov diffusion,
which would manifest itself in different values of $\alpha'$.

\subsection{Non-universality of the energy dependence.}

To study the non universality of $\alpha_0 $ it is necessary to study
the energy dependence of the electroproduction of vector mesons as a
function of $Q^2$. Up to now only two results are known, $\alpha_0
\sim 1.08$ from the $\rho$ meson photoproduction~\cite{photorho}, and
$\alpha_0 \sim 1.30$ as estimated from preliminary HERA data at $Q^2
\sim 10$~GeV$^2$~\cite{ZEUSb}.  The key question is at what $Q^2$
a significant rise of $\alpha_0 $ starts -- this will give a direct
information on the transition region from soft to hard physics.
Optimists of PQCD suggest that the rise may occur already at $Q^2 \sim
3$~GeV$^2$. The same question applies for production of heavier
$\phi$ and $J/\Psi$ mesons. Since the $J/\Psi$ meson is a small object
one may speculate that in this case the rise could start already for
photoproduction (the experimental data indicate that the slope of the
$J/\Psi$ exclusive photoproduction cross section is close to the value
given by the two-gluon form factor of the nucleon). The practical
problem for a quantitative analysis is that no accurate
data on {\bf exclusive} $J/\Psi$ photoproduction at fixed target
energies are available at the moment. Inclusive fixed target data
where the $J/\psi$ meson carries practically the whole momentum of the
projectile photon which are used to extract the exclusive channel seem
to be significantly contaminated by the contribution of the reaction
$\gamma + p \rightarrow J/\Psi + X$ which is peaked at $x_F \equiv
p_v/p_{\gamma}$ close to 1.

\subsection{Non-universality of the t-dependence.}

The slope of the effective Pomeron trajectory $\alpha'$ should
decrease with increasing $Q^2$. This is because the Gribov diffusion
in the impact parameter space, which leads to finite
$\alpha'$~\cite{Gribovdif}, becomes inessential in the hard regime.
This is a consequence of the increase with energy of the typical
transverse momenta of partons. Thus for the reactions $\gamma^* +
N\rightarrow V + N$ the effective $\alpha'$ should decrease with
increasing $Q^2$ while a universal Pomeron exchange approximation
predicts for the energy dependence of the slope
\begin{equation}
B(s)=B(s_0) + 2 \alpha' \ln \left( {s\over s_0} \right)
\end{equation}
with $\alpha' \sim 0.25$~GeV$^{-2}$.

It is possible to look for this effect by comparing the HERA and the
NMC data on the $\rho$ meson production. The universal Pomeron model
predicts that the slope should change from $B \sim
4$~GeV$^{-2}$~\cite{NMC1} to $B \sim 6$~GeV$^{-2}$ at HERA
energies while in the perturbative domain a much weaker change of the
slope is expected.

The slope of the effective Pomeron trajectory $\alpha'$ may depend on
the flavor. It should decrease with the mass of flavor.  Thus it would
be very important to measure the effective $\alpha'$ for the
diffractive photoproduction of $\rho, ~\phi$ and $J/\Psi$.  If PQCD is
important for $J/\Psi$ photoproduction one would expect a smaller
increase of the slope with energy in this case.

\subsection{Non-universality of the gap survival probability.}

The presence of configurations of different size in hadrons (photons)
should also manifest itself in the non universality of the gap survival
probability in the two jet events.  Since the probability of gap
survival is determined by the intensity of the {\it soft} interaction
of the projectile with the target, the survival probability should
increase with increase of $Q^2$, and at fixed $Q^2$ it should be
larger for the heavy $q \bar q$ components of the photon.  Also, the
gap survival probability in the photon case should be substantially
larger than that observed in $p \bar p$ collisions at
FNAL collider~\cite{gapfnal}. This reflects the difference between
$\sigma_{tot}(p \bar p) \approx 80$ mb and the effective cross section
for the interaction of the hadronic components of $\gamma (\gamma^*)$
with nucleon of $\le 30$ mb.

Observation of non-universalities discussed here will shed light on the
structure of the effective Pomeron operating in strong interactions
and will help to address the question about the major source of the
increase of the total cross section of $p \bar p$ interaction --- soft
physics or hard physics of small size configurations.

\subsection{Non-universality of diffraction dissociation}

Since the object which couples to the nucleon in the hard coherent processes
is different from soft Pomeron one may expect a difference between  the value
of the ratio $\left. {{d \sigma^{\gamma^* +p \rightarrow \rho +X} \over dt}
\over
{d \sigma^{\gamma^* +p \rightarrow \rho +p} \over dt}}\right|_{t=0}$ and
similar ratio for soft processes. Qualitatively, one may expect that since the
coupling of
effective Pomeron in hard processes
is more local the ratio of diffraction dissociation and elastic cross sections
should  be substantially smaller  for hard  processes, at least for small
excitation masses.

\section{Summary.}

We have demonstrated that color coherent phenomena should play in QCD
a rather direct role both in the properties of hadrons and in the high
energy collisions. It seems now that recent experimental data confirm
some of the rather nontrivial predictions of QCD and help to elucidate
such old problems as the origin of the Pomeron pole and the Pomeron
cuts in the Reggeon Calculus. Thus we expect that the investigation of
coherent hard and soft diffractive processes may be the key in
obtaining a three dimensional image of hadrons, in helping to search
for new forms of hadron matter at accelerators and in understanding
the problem of inter-nucleon forces in nuclei. Forthcoming high
luminosity studies of diffraction at HERA which will include among
other things the detection of the diffracting nucleon and the $\sigma_L -
\sigma_T$ separation would greatly help in these studies.

 \section*{Acknowledgments}

 We would like to thank J. Bartels, J. Bjorken, S. Brodsky, W.
 Buchmuller, A. Caldwell, J. Collins, J. Ellis, G. Knees, H. Kowalski,
 L. Lipatov, A. Mueller, G. Wolf for the fruitful discussions of the
 interplay of soft and hard physics and of methods of their
 investigation.  Many thanks go to R. Klanner for a careful reading of
 the text.  We would also like to acknowledge the hospitality of DESY
 where part of this work was completed.

\end{document}